\documentclass[a4paper,fleqn]{cas-dc}

\usepackage[authoryear,longnamesfirst]{natbib}

\usepackage{amssymb,amsthm}
\usepackage{graphicx} 
\usepackage{epsfig} 
\usepackage{subcaption}
\usepackage{amsmath} 
\usepackage{bm}

\def\tsc#1{\csdef{#1}{\textsc{\lowercase{#1}}\xspace}}
\tsc{WGM}
\tsc{QE}

\def\x{{\bf x}}
\def\u{{\bf u}}

\def\s{{\bf s}}
\def\z{{\bf z}}
\def\U{{\bf U}}
\def\X{{\bf X}}
\def\q{{\bf q}}
\def\d{{\bf d}}
\def\f{{\bf f}}

\def\0{{\bf 0}}
\def\B{{\cal B}}
\def\Q{{\bf Q}}
\def\A{{\bf A}}
\def\B{{\bf B}}
\def\C{{\bf C}}
\def\D{{\bf D}}
\def\F{{\bf F}}
\def\I{{\bf I}}

\def\H{{\bf H}}

\def\R{{\bf R}}
\def\L{{\bf L}}
\def\W{{\bf W}}
\def\G{{\bf G}}
\def\diag{{\rm diag}}

\def\T{\mathsf{T}}
\def\E{\mathbb{R}}

\begin{document}
\let\WriteBookmarks\relax
\def\floatpagepagefraction{1}
\def\textpagefraction{.001}

\shorttitle{Multi-gated perimeter flow control for monocentric cities}    

\shortauthors{Jusoh and Ampountolas}  

\title[mode=title]{Multi-gated perimeter flow control for monocentric cities:\\ Efficiency and equity}  

%

%

\author[1]{Ruzanna Mat Jusoh}[]

%

\ead{ruzanna@upnm.edu.my}

\affiliation[1]{organization={Department of Defence Science, National Defence University of Malaysia},
            city={57000 Kuala Lumpur},
            country={Malaysia}}

\author[2]{Konstantinos Ampountolas}[orcid = 0000-0001-6517-8369]


\ead{k.ampountolas@uth.gr}



\affiliation[2]{organization={Automatic Control \& Autonomous Systems Laboratory, Department of Mechanical Engineering, University of Thessaly},
            city={38334 Volos},
            country={Greece}}

\cormark[2]

\cortext[2]{Corresponding author}



\begin{abstract}
A control scheme for the multi-gated perimeter traffic flow control problem of monocentric cities is presented. The proposed scheme determines feasible and optimally distributed input flows for various gates located at the periphery of a protected network area. A parsimonious model is employed to describe the traffic dynamics of the protected network. To describe traffic dynamics outside of the protected area,  the basic state-space model is augmented with additional state variables to account for vehicle queues at store-and-forward origin links at the periphery. The multi-gated perimeter flow control problem is formulated as a convex optimisation problem with finite horizon, and constrained control and state variables. This scheme aims to equalise the relative queues at origin links and to maintain the vehicle accumulation in the protected network around a desired set point, while the system's throughput is maximised. For real-time control, the optimal control problem is embedded in a rolling-horizon scheme using the current state of the whole system as the initial state as well as predicted demand flows at origin/entrance links. Furthermore, practical flow allocation policies for single-region perimeter control strategies without explicitly considering entrance link dynamics are presented. These policies allocate a global perimeter-ordered flow to a number of candidate gates at the periphery of a protected network area by taking into account the different geometric characteristics of origin links. The proposed flow allocation policies are then benchmarked against the multi-gated perimeter flow control.  A meticulous study is carried out for a 2.5 square mile protected network area of San Francisco, CA, including fifteen gates of different geometric characteristics. The results showed that the proposed approach is able to manage excessive queues outside of the protected network area and to optimally distribute the input flows, which confirms its efficiency and equity properties. Similar policies are expected to be utilised for dynamic routing and road pricing.
\end{abstract}




\maketitle

\section{Introduction}
Traffic congestion on urban road networks is deemed inefficient due to road operations and excessive traffic demand, which calls for drastic solutions. Traffic management via traffic lights has been the subject of intensive research for the past six decades to improve urban mobility and relieve congestion \citep{PDDKW:03}. Although many advanced traffic signal control strategies have been developed to tackle congestion in urban areas, still a challenge is the deployment of efficient traffic management schemes for preventing overcrowding in central areas of monocentric cities or polycentric megacities. 

The problem of preventing overcrowding in central areas (due to traffic congestion or air pollution) has been traditionally addressed with the deployment of congestion-pricing schemes that is based on historical information or  short-term traffic prediction. Nobel Laureate William Vickrey was perhaps the first who advanced the idea of congestion-pricing during the 1960s \citep{Vickrey63,Vickrey69}. Application of this idea includes the congestion-pricing or $\rm CO_2$ emissions-based charging scheme for the London central area and similar schemes for the central Singapore area, central neighbourhoods of Stockholm, Zurich, and a number of cities in the US (see e.g., \citet{Litman2005}). Congestion pricing schemes range from charging a constant congestion or daily fee to dynamic charging that is based on a pay-as-you-use principle. E-ZPass or similar automatic billing systems have been widely used to realize automatic charging in a number of cities in the US, including New York, Pennsylvania, Chicago, Los Angeles and San Francisco; particularly to allow drivers to pay in order to use express lanes that were previously available only to high-occupancy vehicles. 

An alternative to congestion-pricing is a perimeter flow control or gating scheme, where vehicles are delayed or re-routed outside of the periphery of a network area to protect downstream central areas from overcrowding in the sense of limiting the entrance in the network when close to overload. To advance in a systematic way this research the performance of city-wide or neighbour traffic and road infrastructure should be efficiently assessed and short-term predictions of system dynamics should be available in real-time  \citep{Daganzo:07}. The performance of road infrastructure is usually assessed by microscopic models at the link or junction level. In an attempt to assess the performance of urban road networks at a macroscopic level, a parsimonious but not accurate model is often used, which primarily shows the relationship between average network flow and vehicle accumulation or traffic density. This model is the so-called Macroscopic or Network Fundamental Diagram (MFD or NFD) of urban road networks.

\begin{figure}[tb]
\begin{center}
\includegraphics[width=.7\columnwidth]{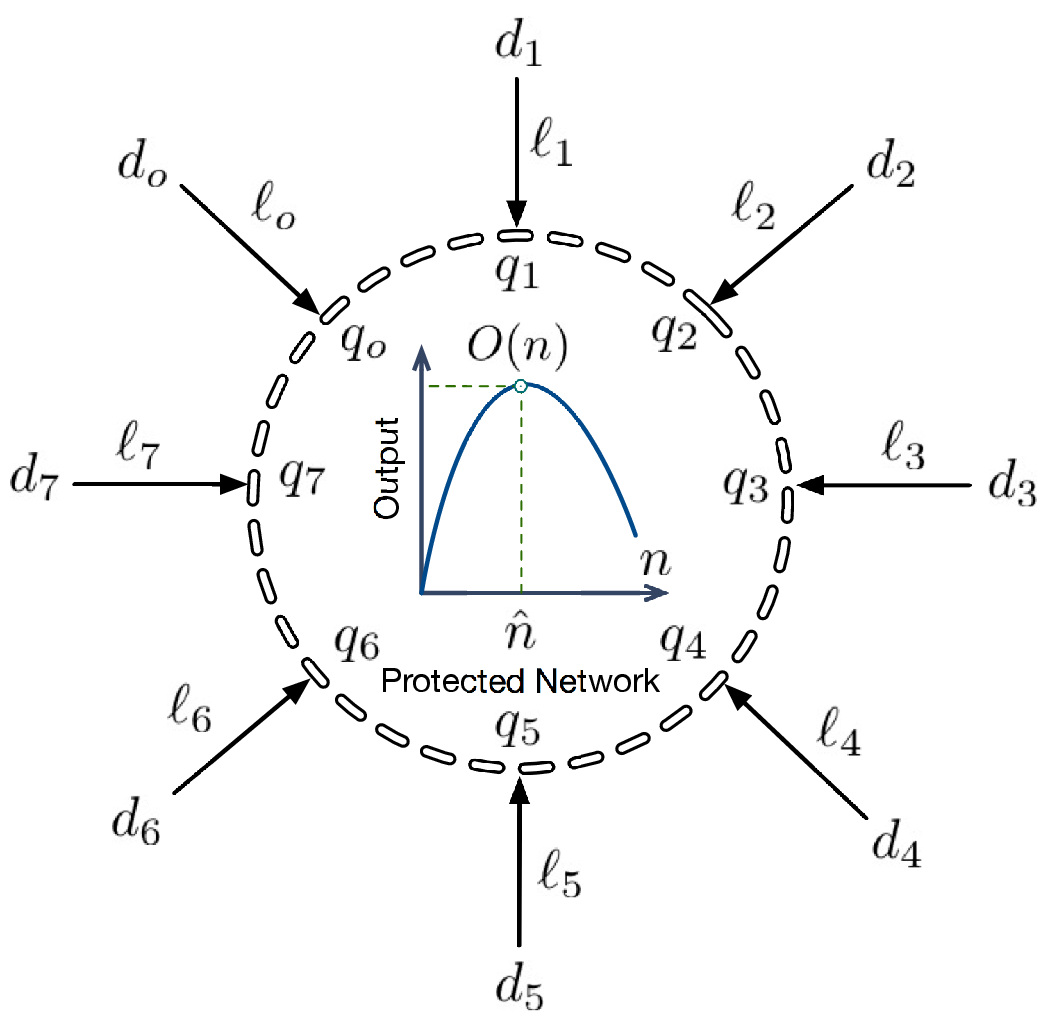}    
\caption{Protected network with entrance link dynamics.} 
\label{fig:PN}
\end{center}
\end{figure}

The main feature of the NFD with a concave like-shape, as shown in Figure \ref{fig:PN}, is that for a (more or less)  critical vehicle accumulation $\hat n$ flow capacity is reached i.e., maximum throughput. This property of the NFD can be utilised to introduce perimeter flow or gating control policies to improve urban mobility and relieve congestion in homogeneous urban areas \citep{Daganzo:07}. A perimeter flow control policy ``meters" the input flow to the system and hold vehicles outside of a protected network (PN) area if necessary, so as to maximise the network throughput. This is usually achieved by controlling traffic lights, toll stations or automatic billing systems located at the periphery, the perimeter of a protected network area, which exhibits a well-defined NFD. 

This paper contributes to the state-of-the-art by:
\begin{itemize}
	\item Describing the traffic dynamics outside of the protected area using an augmented state-space model with additional state variables for the queues at store-and-forward entrance links at the periphery. 

	\item Developing a multi-gated perimeter flow control (MGC) scheme to optimally distribute input flow values (or feasible entrance link green times) to avoid queues and delays at the perimeter of a protected area while overall system's output is maximised. 
			
	\item Employing a model-predictive control scheme to  optimally distribute input flows at the perimeter of the protected network that explicitly considers queue dynamics and constraints at store-and-forward links.
	
	\item Developing practical flow allocation policies for single-region perimeter control strategies without explicitly considering entrance link dynamics. These policies allocate a global perimeter-ordered flow to a number of candidate gates at the periphery of a protected network area by taking into account the different geometric characteristics of origin links. 
\end{itemize}

The rest of the paper is structured as follows: Section \ref{sec:LR} reviews the relevant literature of perimeter control in urban regions with a focus on fairness and equity. Section \ref{sec:IntegratedModel} presents the proposed macroscopic model for the multi-gated perimeter traffic flow control problem. Section \ref{sec:multigated} develops the multi-gated perimeter flow control strategy via constrained rolling horizon optimisation. Practical perimeter-ordered flow allocation policies are presented in Section \ref{sec:allocationpolicies}. Section \ref{sec:CaseStudy} demonstrates the effectiveness of the proposed multi-gated perimeter flow control and allocation policies and provides many insights on their efficiency and equity features. Conclusions are given in Section \ref{sec:conclusions}.

\section{Literature Review}\label{sec:LR}

Performance assessment of city-wide traffic appears to have been the subject of study in the late sixties \citep{Smeed:1966,Godfrey:1969,Zahavi:1972ab}.  In particular, a macroscopic model of steady-state urban traffic was proposed by \citet{Godfrey:1969,HP:1979}, further developed by \citet{ArdHer:1987,Olszewski1995273,Daganzo:07,DaganzoGerol:08,Daganzo:2011,Farhi201185,Mahmassani2013} and fitted to experimental data by \citet{MHWJHR:1987,GerolDaganzo:08,7171036}, among others. This network-wide traffic flow model, the NFD,  assumes (under certain regularity conditions like homogenaity) that traffic flow dynamics of urban areas can be treated macroscopically as a single-region dynamic system with vehicle accumulation (or traffic density) as a state variable. 

Theoretical work based on variational theory, which has also been validated with experimental data to some extent, shows that the shape of the NFD depends on the network topology (length of roads, structure of network, number of lanes, etc.), the traffic signal settings (green splits, cycle length and offsets), and the traffic mode composition in case of mixed multimodal traffic \citep{NGBB:2012,NG_NZ_KAPartC:2014}. The effect of traffic demand and traffic-responsive signal control in systems described by NFD dynamics is discussed in \citet{Aboudolas2010,Haddad20121159,Gayah2011643,sab:2014,Zhang:2013,Gayah2014255}. The shape of the NFD can also be influenced by the degree of spatial heterogeneity in the distribution of congestion as revealed in a number of studies, see e.g., \citet{Mazl:2010,GerolSun11,Daganzo:2011,Knoop:2013}. However, by controlling the periphery of the protected network area with a well-defined NFD and not intervening in the traffic lights inside the network, the shape of the NFD remains about the same; and, thus perimeter control is well-defined and appropriate for preventing overcrowding \citep{MARIOTTE2017245}.

The concave shape of the NFD can be used to introduce perimeter flow policies to improve mobility and mitigate congestion in monocentric homogeneous urban areas \citep{Mehdi:2012a,Mehdi:2013,Haddad2014315} or inhomogeneous polycentric cities with multiple centres of congestion, if these cities can be clustered into a small number of homogeneous clusters \citep{Ji20121639,Saeedmanesh2016250,Venkatasubramaniam2023}. Similarly, perimeter flow control policies can be developed for polycentric heterogeneous cities \citep{GerolHaddad:2013,KA_NG_PartB:2013,Kouvelas201726,Haddad2017495,Haddad20171}.  

Recent studies extend the aforementioned macroscopic modelling approach from single-modal to bi-modal networks, where cars and buses share the same infrastructure and interact \citep{NG_NZ_KAPartC:2014, Chiabaut2014, Chiabaut2015}. In these works, both bi-modal vehicular dynamics and passenger flow dynamics are of interest, given that buses carry more passengers. The existence of a mixed traffic, bi-modal (three-dimensional) NFD is investigated in \citet{NG_NZ_KAPartC:2014} via a meticulous micro-simulation study for a large network with dynamic features. The extended versions of the NFD, so-called 3D-vMFD and 3D-pMFD, relate the accumulation of cars and buses to vehicular and passenger flows, respectively. Furthermore, provide inspiration for investigating perimeter flow control strategies for mixed bi-modal networks as in \citet{Ampountolas:2017}. Empirical observations besides simulation studies are currently under development.

The equity properties of perimeter flow control have not attracted considerable attention in the literature, although it is an important characteristic of any potential practical perimeter flow control application. Recent research efforts on equity properties of perimeter flow control have focused on developing approaches that prioritize fairness and equity while optimizing network efficiency, see e.g., \cite{RMJ2017,KEYVANEKBATANI2021104762,MOSHAHEDI2023,HOSSEINZADEH2023}.  Fairness may be assessed through regional speed while equity can be quantified by the total travel time \citep{MOSHAHEDI2023} or delays \citep{KEYVANEKBATANI2021104762}. Route guidance control schemes can also consider fairness to maximize the proportion of travellers' utilities and prevent the network from becoming congested while also ensure fairness in the distribution of network resources. Compliance of drivers to routing instructions, enhance fairness between different paths in the network and improve network traffic conditions \citep{HOSSEINZADEH2023}. 

Except of a few works (see e.g., \citet{Csikos2015, RMJ2017b,Haddad20171,NI2019358,NI2020164,SIRMATEL2021104750,MERCADER2021104718,KEYVANEKBATANI2021104762,GUO2021125401}), studies on the perimeter flow control assume that a single input flow ordered by a perimeter control strategy should be distributed in a non-optimal way (e.g., equal distribution) to a number of candidate junctions at the periphery of the network, i.e., without taking into account the different geometric characteristics of origin links such as length, storage capacity, etc. A distribution policy applied independently to multiple gates of a protected network area would be efficient in case of unconstrained origin link queues for vehicle storage (i.e., infinite storage capacity). However, gated queues at origin links must be restricted to avoid interference with adjacent street traffic outside of the protected network area and geometric characteristics of the different gates must be taken into account in the optimisation. Thus, limited origin links storage capacity/geometric characteristics and the requirement of equity for drivers using different gates to enter a protected area are the main reasons towards multi-gated perimeter flow control, in which the current paper contributes.

\section{Macroscopic modelling of monocentric cities with entrance link dynamics}\label{sec:IntegratedModel}

\subsection{Modelling}
Assume there exists a well-defined function $O[n(t)]$ (veh/h) of the vehicle accumulation $n(t)$, $t \ge 0$, which provides the estimated rate flow (output) at which vehicles complete trips per unit of time either because they finish their trip within a monocentric network area or because they move outside of the network. This function, known as NFD, describes steady-state behaviour of monocentric homogeneous cities if the input to output dynamics are not instantaneous and any delays are comparable with the average travel time across the network \citep{Daganzo:07}.  The output (throughput) function $O[n(t)]$ of a network can be easily determined if trip completion rates or Origin-Destination (OD) data are available in real-time, e.g., from vehicles equipped with GPS trackers capable of providing locational data at any given time. Alternatively, the output  can be expressed as $O[n(t)] \triangleq (l/L) O_c[n(t)]$, where $L$ (km) is the average trip length in the network; $l$ (km) is the average link length; and, $O_c$ (veh/h) is the total network circulating flow, as reflected by the observations of a number of (mid-block) inductive loop detectors (or other sensors) placed at appropriate network locations. In general, the circulating flow $O_c$ can be estimated by Edie's generalised definition of flow if $n(t)$ is observed in real-time, i.e.\ weighted average of link flows with link lengths \citep{Edie:1963}.


\begin{table}[tb]
\caption{Notation.}\label{tbl:notation}\centering
 \begin{tabular}{ll}\hline
Notation         & Meaning \\   \hline 
$n$             & Vehicle accumulation\\
$n_{\max}$             & Maximum vehicle accumulation at the PN\\
$O(n)$ & Network fundamental diagram\\
$\cal O$ & Set of origin links at the periphery of the PN\\
$d_o$            & Traffic demand at controlled gate $o \in \cal O$\\
$\d$            & Disturbance vector \\
$\hat{\d}$      & Set point for disturbance variables\\
$\kappa$ & Discrete time index\\
$\ell_o$            & Vehicle queue at origin link $o \in \cal O$\\
$\ell_{o,\min}$            & Min capacity at origin link $o\in \cal O$\\
$\ell_{o,\max}$            & Max capacity  at origin link $o\in \cal O$\\
$d_{n}$              & Uncontrolled traffic demand within the PN\\
$q_{G}$& Global perimeter-ordered flow\\
$q_{o}$             & Input flow at controlled gate $o\in \cal O$\\
$q_{o,\min}$             & Min permissible outflows for origin link $o\in \cal O$\\
$q_{o,\max}$             & Max permissible outflows for origin link $o\in \cal O$\\
$q_{{\rm out}} (t)$ & Total outflow of the PN at time $t$  \\
$q_o$            & Ordered input flow at controlled gate $o\in \cal O$\\
$\hat{q}$           & Nominal input flow at controlled gate $o\in \cal O$\\
$r_{o}$ & Capacity ratio of each entrance link $o\in \cal O$\\
$\u$            & Input/control vector\\
$\hat{\u}$      & Set point for input/control variables\\
$\x$            & State vector\\
$\hat{\x}$      & Set point for state variables\\ 
$N_o$& Optimisation horizon\\
$N_p$& Prediction horizon\\
$\A$ ($\tilde{\A}$) & State matrix (augmented matrix)\\
$\B$ ($\tilde{\B}$) & Control matrix (augmented matrix)\\
$\C$ ($\tilde{\C}$) & Disturbance matrix (augmented matrix)\\
\hline 
\end{tabular}
\end{table}

Consider now a protected network area, which exhibits a well-defined NFD, with a number of origin links (or controlled gates) $o \in {\cal O} = \{1, 2, \ldots\}$ located at its periphery, as shown in Figure \ref{fig:PN}. The set $\cal O$ includes all the origin links whose outflow is essentially entering into the protected network from a number of controlled gates/entrances (e.g.\ signalised junctions or toll stations).  In principle, the origin links at the periphery of the protected network would have different geometric characteristics, i.e., length, number of lanes, capacity, saturation flows. Let $q_{o} (t)$ (veh/h) be the outflow of gate $o \in {\cal O}$ at time $t$. Also, let $q_{{\rm out}} (t)$ (veh/h) and $d_n(t)$ (veh/h) be the total outflow and the uncontrolled traffic demand (disturbances)  of the protected network at time $t$, respectively. Note that $d_n(t)$ includes both internal (off-street parking for taxis and pockets for private vehicles) and external (from the periphery) non-controlled inflows. The dynamics of the system are governed by the following nonlinear conservation equation:
\begin{equation}\label{eq:PNd}
\dot{n}(t) =  \sum_{o = 1}^{|{\cal O}|}q_o(t-\tau_o) - \min\{\beta_0, q_{\rm out} (t)\} + d_n(t),
\end{equation}
where $\beta_0$ is a maximum exit flow to limit the outflow of the protected network under heavy congestion,  $q_{\rm out} (t)$ is in general a nonlinear function of vehicle accumulation $n(t)$, and $\tau_o$ is the travel time needed for vehicles to approach the protected network area from origin link $o \in {\cal O}$.  The time lags $\tau_o$ may be translated into an according number of time steps for a discrete-time representation, provided a closed system with inflows $q_o$ and outflow $q_{\rm out}$. Without loss of generality, we assume that $\tau_o = 0$, $\forall\ o \in {\cal O}$, i.e., vehicles released from the controlled gates can immediately get access to the protected network. This assumption will be removed later in Section \ref{sec:delays} through a standard recasting process. Moreover, since the system evolves slowly with time $t$, we may assume that outflow $q_{\rm out}(t) \propto O_c(n(t))$, and it may thus be given in terms of the output $O[n(t)]$. Note that $q_o(t)$, $o \in \cal O$ are the input variables of the controlled gates/entrances, to be calculated by a mutli-gated perimeter flow control strategy.

To describe traffic dynamics outside of the protected area, we augment the basic state-space model \eqref{eq:PNd} with additional state variables for the queues at store-and-forward entrance links at the periphery. Each origin link $o$ receives traffic demand $d_o$ and forward it into the protected network, as shown in Figure \ref{fig:PN}. The queuing model for the entrance link dynamics is described by the following conservation equation:
\begin{equation}\label{eq:Qd}
\dot{\ell_o}(t) = d_o(t) - q_o(t), \, o \in {\cal O} = \{1, 2, \ldots\},
\end{equation}
where $\ell_o(t)$ (veh) and $d_o(t)$ (veh/h) are the vehicle queue and traffic demand in origin link $o$ at time $t$, respectively. 

The integrated model  \eqref{eq:PNd}--\eqref{eq:Qd} can be extended to consider a broader class of state and control constraints. For example, inequality state and control constraints may be introduced to preserve congested phenomena within the protected network and to avoid long queues and delays at the perimeter of the network where gating is literally applied. These constraints may be brought to the form,
\begin{equation}\label{eq:constraints}
\begin{aligned}
& 0 \le n(t) \le n_{\max},\\ 
& 0 \le \ell_o(t) \le \ell_{o, \max}, \, o \in {\cal O}  = \{1, 2, \ldots\},\\
& q_{o, \min} \le q_o(t) \le q_{o, \max}, \, o \in {\cal O}  = \{1, 2, \ldots\},
\end{aligned}
\end{equation}
where $n_{\max}$ is the maximum vehicle accumulation of the protected network; $\ell_{o, \max}$ is the maximum permissible capacity of link $o \in {\cal O}$; $q_{o, \min}$, $q_{o, \max}$ are the minimum and maximum permissible outflows, respectively; and, $q_{o, \min} > 0$ to avoid long queues and delays at the periphery of the network. Link capacities and maximum vehicle accumulation depend on geometric characteristics of the origin links (length, number of lanes) and the topology of the protected network, respectively. Minimum and maximum permissible outflows can easily be determined given saturation flows, minimum and maximum green times, and cycle times of a nominal traffic signal plan (or corresponding toll ticket) at each controlled gate of the protected network.

State equation \eqref{eq:Qd} is linear, though a more accurate nonlinear form can be written to account storage capacity and dispersion of the flow phenomena within store-and-forward origin links. In this case, the outflow function of each gate $o \in {\cal O}$ is given by: 
\begin{equation}\label{eq:Realoutflow}
q_o(t) = \begin{cases}
q_{o, \min}, 					& \text{if }\, n(t) \ge c n_{\max} \\
\min\{d_o(t), \tilde{q}_o(t),q_{o, \max}\}, 		& \text{otherwise}
\end{cases},
\end{equation} 
where $\tilde{q}_o(t)$, $o \in {\cal O}$, are now the input variables to be calculated by a multi-gated perimeter flow control strategy, $c \in (0.5, 1)$ is a scalar introduced to prevent overflow phenomena within the protected network area, and $q_{o, \max}$ is the maximum permissible outflow under a nominal traffic signal plan (or toll ticket) at controlled gate $o \in {\cal O}$. Note that, when using \eqref{eq:Realoutflow}, the state constraints in \eqref{eq:constraints} for all origin links $\ell_{o}$, $o\in \cal O$ are considered indirectly and may hence be dropped; indeed the gated outflow in  \eqref{eq:Realoutflow} becomes zero if there is no queue in the corresponding origin link or becomes minimum (provided a minimum green time at the controlled gates) if the protected network is oversaturated (determined by $c$).

The presented model can be viewed as a nonlinear process with input variables $\u^\T \triangleq \begin{bmatrix}q_1 & q_2 & \cdots & q_{|{\cal O}|}\end{bmatrix}$, state variables $\x^\T \triangleq \begin{bmatrix}n & \ell_1 & \ell_2 & \cdots & \ell_{|{\cal O}|}\end{bmatrix}$, and disturbances $\d^\T \triangleq \begin{bmatrix}d_n & d_1 & d_2 & \cdots & d_{|{\cal O}|}\end{bmatrix}$. Then, the continuous-time nonlinear state system \eqref{eq:PNd}, \eqref{eq:Qd}, \eqref{eq:Realoutflow} with constraints \eqref{eq:constraints} for a protected network with controlled gates $o \in {\cal O}$, may be rewritten in compact vector form as, 
\begin{align}\label{eq:nonlinear}
& \dot{\x}(t) = \f\left[\x(t),\u(t),\d(t),t\right], \, t \ge 0, \quad \x(0) = \x_0,\\ \label{eq:c1}
& \0 \le \x(t) \le \x_{\max},\\ \label{eq:c2}
& \u_{\min} \le \u(t) \le \u_{\max},
\end{align}
where $\f$ is a nonlinear vector function reflecting the right-hand side of \eqref{eq:PNd}--\eqref{eq:Qd}; $\x_0$ is a known initial state; and $\x_{\max}$, $\u_{\min}$, $\u_{\max}$ are vectors of appropriate dimension reflecting the upper and lower bounds of constraints \eqref{eq:constraints}. 

Assuming a nonlinear representation of $q_{\rm out} (t)  \triangleq O[n(t)]$, the continuous-time nonlinear model \eqref{eq:nonlinear} may be linearised around some set point $\hat{\s}^\T \triangleq \begin{bmatrix}\hat{\x} & \hat{\u} & \hat{\d}\end{bmatrix}$, and directly translated into discrete-time, using Euler first-order time discretisation with sample time $T$, as follows:
\begin{equation}\label{eq:linear_v_dt}
\Delta \x(k+1) =  \A \Delta \x(k) + \B \Delta \u(k) + \C\Delta \d(k),
\end{equation}
where $k  = 0, 1, \ldots, N_o-1$ is a discrete time index with optimisation horizon $N_o$; $\Delta (\cdot) \triangleq (\cdot) - \hat{\cdot}$ for all vectors; and $\A = \partial \f / \partial 
\x|_{\hat{\s}}$, $\B = \partial \f / \partial \u|_{\hat{\s}}$, $\C = \partial \f / \partial  \d|_{\hat{\s}}$ are the state, control, and disturbance matrices, respectively. This discrete-time linear model is completely controllable and reachable, and will be used as a basis for control design.

The sample time interval $T$ is literally selected to be a common multiple of cycle lengths of all controlled gates at the periphery of the protected network, while $T \in [3, 5]$ minutes is usually appropriate for constructing a well-defined outflow function $O[n(t)]$, given experimental data. In principle, origin link dynamics \eqref{eq:Qd} are much faster than the dynamics of the protected network \eqref{eq:PNd} (governed by the network fundamental diagram, which evolves slowly in time). Therefore two different time steps $T_n$ and $T_{\ell}$ (where $T_{\ell} \ll T_{n}$) can be employed for \eqref{eq:PNd} and \eqref{eq:Qd}, respectively, to account storage capacity and dispersion of the flow phenomena within store-and-forward origin links; and thus increase model accuracy. By introducing different time steps, the state variables for origin links $\ell_o$, $o\in {\cal O}$, are allowed to change their value more frequently than the state of the protected network $n$ and control variables $q_o$, $o\in {\cal O}$.

\subsection{Dealing with time delays}\label{sec:delays}

Standard methods in control are not directly applicable to problems involving time delays in controls, like the delay times $\tau_o$ in \eqref{eq:PNd}. This difficulty can be readily eliminated by introducing additional auxiliary state variables $z_o$. For example, for a control variable $q_{o}(k-\kappa_o)$ appearing in the discrete-time model \eqref{eq:linear_v_dt}, where $\kappa_o$ (integer) is the corresponding number of discrete time steps for delay time $\tau_o$, one may introduce $\kappa_o$ auxiliary state equations as follows
\begin{equation}
\begin{aligned}
z_1(k+1) &= q_{o}(k)\\
z_2(k+1) &= z_1(k)\\
&\vdots\\
z_{\kappa_o}(k+1) &= z_{\kappa_o-1}(k)
\end{aligned}
\end{equation}
and substitute $z_{\kappa_o}(k)$ in all model equations where $q_{o}(k-\kappa_o)$ appears. The augmented discrete-time model can be written as,
\begin{equation}\label{eq:linear_augm}
\Delta \tilde{\x}(k+1) =  \tilde{\A} \Delta \tilde{\x}(k) + \tilde{\B} \Delta \tilde{\u}(k) + \tilde{\C}\Delta \d(k),
\end{equation}
where $\tilde{\x}(k) = \left[ \x(k) \quad \z(k)\right]^\T$ is the augmented state vector, and $\tilde{\A}, \tilde{\B}, \tilde{\C}$ are the augmented state, control, and demand matrices, respectively. The augmented linear model \eqref{eq:linear_augm} (instead of \eqref{eq:linear_v_dt}) can be then used as a basis for control design if necessary.

\section{Multi-gated perimeter flow control (MGC) via constrained rolling horizon control}\label{sec:multigated}

A natural control objective for the traffic system considered is to minimise the total time that vehicles spend in the system including both time waiting at origin links to enter and time traveling in the protected network. Actually the minimisation of the total time spent is equivalent to the maximisation of the total exit flow (or trip completion rate) from the protected network, under the assumption of given control-independent demand inflows and of infinitely long origin link queues (unconstrained vehicle storage). However, such a policy may induce unbalanced gating of vehicles at the origin links of the protected network, and, as a consequence, may lead to long queues and overflow phenomena within origin links. Unbalanced gating would also violate the requirement of equity for drivers using different gates to enter a protected network area. 

Given these observations, a suitable control objective for a protected network area with origin links queue dynamics aims at: (a) equalising the relative vehicle queues $\ell_o/ \ell_{o, \max}$, $o \in \cal O$ over time, and (b) maintaining the vehicle accumulation in the protected network around a set (desired) point $\hat{n}$ while the system's throughput is maximised. A quadratic criterion that considers this control objective has the form
\begin{equation}\label{eq:CF_LQ}
J = \frac{1}{2} \sum_{k=0}^{N_o-1} \Big(\|\Delta\x(k)\|^2_{\bf Q} + \|\Delta \u(k)\|^2_{\bf R}\Big),
\end{equation}
where $\Q$ and $\R$ are positive semi-definite and positive definite diagonal weighting matrices, respectively. The diagonal elements of $\Q$ (see definition of vector $\x$) are responsible for balancing the relative vehicle accumulation of the protected network $n/n_{\max}$ and the relative vehicle queues $\ell_o/\ell_{o,\max}$, $o \in \cal O$. Given that vehicle storage in the protected network is significantly higher than in the origin links, a meticulous selection of diagonal elements is required. A practicable choice is to set $\Q = \diag(1/w, 1/\ell_{1,\max}, \ldots, 1/\ell_{|{\cal O}|,\max})$, where the scale of $w \ll n_{\max}$ is of the order of $\sum_{o=1}^{|{\cal O}|}\ell_{o,\max}$ to achieve equity. It becomes quite clear here that equity at origin links and efficiency of the protected network area are partially competitive criteria, hence a perimeter flow control strategy should be flexible enough to accommodate a particular trade-off (i.e. to give priority to the protected network or the outside area, e.g.\ to manage better excessive queues) to be decided by the responsible network authorities. Finally, the choice of the weighting matrix $\R \triangleq r\I$, $r > 0$ can influence the magnitude of the control actions and thus $r$ should be selected via a trial-and-error process. 

Rolling horizon (or model-based predictive) control  is a repetitive optimisation scheme, where at each time step an open-loop optimal control problem with finite horizon $N_o$  and predicted demands $\d(k)$ over a prediction horizon $N_p$ is optimised, then only the first control move is applied to the plant and the procedure is carried out again. This rolling-horizon procedure closes the loop, that is, it avoids myopic control actions while embedding a dynamic open-loop optimisation problem in a responsive environment. Predicted demand flows $\d(k)$ may be calculated by use of historical information or suitable extrapolation methods.

Given the known initial state  $\x(0) = \x_0$, a static convex optimisation problem may be formulated over $N_o$ due to the discrete-time nature of the involved process. To see this, assume $N_o = N_p$ and define the vectors 
\begin{equation}
\begin{aligned}
\Delta\X &\triangleq \begin{bmatrix}\Delta\x(1)^\T & \Delta\x(2)^\T & \cdots & \Delta\x(N_o)^\T\end{bmatrix}^\T\\
\Delta\U &\triangleq \begin{bmatrix}\Delta\u(0)^\T & \Delta\u(1)^\T & \cdots & \Delta\u(N_o-1)^\T\end{bmatrix}^\T\\
\Delta\D &\triangleq \begin{bmatrix}\Delta\d(0)^\T & \Delta\d(1)^\T & \cdots & \Delta\d(N_p-1)^\T\end{bmatrix}^\T. 
\end{aligned}
\end{equation}
Assuming now availability of demand flow predictions at the origin links of the protected network over a prediction horizon $N_p$, i.e.\ $\Delta\d(k) \neq \0$, $k = 0, 1, \ldots, N_p-1$, minimisation of the performance criterion \eqref{eq:CF_LQ} subject to \eqref{eq:linear_v_dt} leads to the analytical solution: 
\begin{equation}\label{eq:QP_without_C}
\Delta \U = - \H^{-1} \F\big[\x(0) + \G\Delta \D\big],
\end{equation}
where $\H \triangleq \bm{\Gamma}^\T{\cal Q}\bm{\Gamma} + {\cal R}$ is the Hessian of the corresponding quadratic program (QP), $\F \triangleq \bm{\Gamma}^\T{\cal Q}\bm{\Omega}$, and $\G = \bm{\Gamma}^\T {\cal Z}$. The matrices $\bm{\Gamma}$ and $\bm{\Omega}$ may be readily specified from the integration of \eqref{eq:linear_v_dt} starting from the initial point $\x(0)$, while ${\cal Q}$, ${\cal R}$, $\cal Z$ are weighting matrices (in function of $\Q$, $\R$, and $\C$) over the optimisation $N_o$ (see e.g., \citet{Goodwin:2010}). Given that $\R \succ \0$ in the cost criterion \eqref{eq:CF_LQ} the Hessian $\H$ is positive definite, and thus the QP is convex and has a global optimum. Note that the third term may be regarded as a feedforward term, accounting for future disturbances. Clearly for $N_o \to \infty$ and vanishing disturbances, i.e., $\Delta\d(k) = \0$, $k = 0, 1, \ldots, N_p-1$, a Linear-Quadratic or a Linear-Quadratic-Integral regulator may be derived as in \citet{KA_NG_PartB:2013}.  

Using the above formalism, we can express the problem of minimising \eqref{eq:CF_LQ} subject to the equality constraints \eqref{eq:linear_v_dt} and inequality constraints \eqref{eq:c1}--\eqref{eq:c2} as follows:
\begin{equation}\label{eq:QP}
\begin{aligned}
 \min_{\U} &\quad \frac{1}{2} \U^\T \H \U + \U^\T \big[\F\x(0) -\H\hat{\U}+ \G\Delta \D\big]\\ 
 \text{subject to:} & \quad \L \U \le \W
\end{aligned}
\end{equation}
where $\L$ and $\W$ are matrices reflecting the lower and upper bounds of the state and control constraints (given state integration  starting from the initial point $\x_0$) over the optimisation horizon $N_o$ (see e.g.\ \citet{Goodwin:2010}).  Once the open-loop QP problem \eqref{eq:QP} is solved from the known initial $\x(0) = \x_0$ and predicted disturbances $\d(k), k = 0, 1, \ldots, N_p-1$, the rolling horizon scheme applies, at the current time $k$, only the first control move, formed by the first $m$ components of the optimal vector $\U^\ast(\x_0)$ in \eqref{eq:QP} given by: 
\begin{equation}\label{eq:QP_opt}
\U^\ast(\x_0) = \text{arg}\min_{\L\U \le \W}\frac{1}{2} \U^\T \H \U + \U^\T \big[\F\x(0) + \G\Delta \D\big].
\end{equation}
This yields a control law of the form,
\begin{equation}\label{eq:QP_with_C}
\u(k) = {\cal M}\big[\x(k),\d(\kappa)\big],\,  \kappa = k, k+1, \ldots, k+N_p-1,
\end{equation}
where $\x(k) = \x_0$, $k = 0, \ldots, N_o-1$ is the current state of the system and ${\cal M}$ is a control policy from the state and disturbance spaces to control. Then the whole procedure is repeated at the next time instant, with the optimisation horizon kept constant. Note that the analytical solution \eqref{eq:QP_without_C} for the unconstrained problem is of particular interest; given that the optimal solution for the constrained problem has a similar form in a region of the state space where the state of the system vanishes, i.e. $\Delta \x = \0$ or $\x = \hat{\x}$. 

\section{Practical perimeter-ordered flow allocation policies}\label{sec:allocationpolicies}

\subsection{Motivation}

This section presents practical flow allocation policies for single-region perimeter control strategies without explicitly considering entrance link dynamics.  The problem under consideration is to allocate a global perimeter-ordered flow to a number of candidate gates/junctions at the periphery of the network by taking into account the different geometric characteristics of origin links, i.e., length, number of lanes, storage capacity, etc. The global flow $q_{\rm G}(k) = \sum_{o = 1}^{|{\cal O}|}q_o(k)$ at discrete time $k$ can be ordered by any perimeter control strategy. For instance, a single-input single-output controller with only state equation \eqref{eq:PNd} and cost criterion \eqref{eq:CF_LQ}, or the bang-bang policy proposed by \citet{Daganzo:07}, or the feedback controllers proposed by \citet{Mehdi:2012a,KA_NG_PartB:2013}, or other similar strategies. Previous perimeter flow control strategies without explicitly considering entrance link dynamics (e.g.\ \citet{Mehdi:2012a,KA_NG_PartB:2013,Ampountolas:2017}), assume that a single-ordered input flow is  distributed in a non-optimal or a posteriori way (e.g., equally distributed or with respect to saturation flows) to a number of candidate gates at the periphery of a protected networks area. These strategies employ a two-step procedure. Firstly, an ordered flow is obtained from an unconstrained controller that does not  directly incorporate the operational constraints into the controller synthesis. Secondly, the ordered-flow is then distributed to equivalent entrance link green stages at the perimeter with the help of a flow allocation policy. In the sequel, we propose two perimeter-ordered flow allocation policies to facilitate the real-time deployment of such strategies. These flow allocation policies are later benchmarked against the proposed multi-gated perimeter flow control in Section \ref{sec:CompAlloc}.  

\subsection{Capacity-based flow Allocation Policy (CAP)}

In principle, the distribution of the global perimeter-ordered flow among the controlled gates should be according to appropriately predefined portions. These portions are typically proportional to the links' vehicle storage capacities or nominal flows. In other words, links with high storage capacity or high nominal flows will carry more flow. Moreover, drivers waiting in gated links with similar storage capacity to enter the network from the periphery would potentially experience similar delays. In this section, we propose a capacity-based allocation policy, given that storage vehicle capacity is more important than the nominal saturation flow of each link, particularly under strong gating. In a similar vein, one can develop an allocation policy based on the nominal flows.       

The proposed Capacity-based flow Allocation Policy (CAP), distributes the prevailing global perimeter-ordered flow $q_{\rm G}(k)$ to the controlled gates  $o \in {\cal O} = \{1, 2, \ldots\}$ according to the link storage capacities $\ell_{o, \max}$. The desired distribution attempted via CAP is meant to be active both during the emptying and gating phases, so that equal free relative vehicle storages are provided to each entrance link (in spite of different storage capacities) in the event of strong gating or upstream traffic demand. To start with, the capacity ratio of each entrance link is defined by:   
\begin{align}  \label{eq:ratioCap}
r_o = \frac {\ell_{o, \max}}{\sum_{i = 1}^{|\cal O|} \ell_{i, \max}}, \quad o = 1, 2, \ldots, {|\cal O|}.
\end{align}
Given the global prevailing perimeter-ordered flow $q_{\rm G}(k)$  at discrete time $k$, the individual sub-flows (ordered input flow at controlled gates) are then determined by: 
\begin{equation}
q_{o}(k) = \hat q_{o} +  r_o \times \Bigg(q_{\rm G}(k) - \sum_{i=1}^{|\cal O|} \hat q_{i}\Bigg),  \, o = 1, 2, \ldots, {|\cal O|},
\end{equation}
where $\hat q_{o}$ is the nominal input flow at origin $o \in \cal O$.

\subsection{Optimisation-based flow Allocation Policy (OAP)}

The Optimisation-based flow Allocation Policy (OAP) aims at minimising the relative difference between ordered input flow and nominal flow at each controlled gate. To this end, the following optimisation problem can be formulated and solved at each discrete time $k$, provided a global perimeter-ordered flow $q_{\rm G}(k)$:
\begin{equation}\label{eq:OAP}
\begin{aligned}
 \min_{q_o(k)} &\quad \frac{1}{2} \sum_{o=1}^{|\cal O|} \frac{\big[q_{o}(k) - \hat q_{o}\big]^2}{\hat q_o}\\  
 \text {subject to:} & \quad  \sum_{o=1}^{|\cal O|} q_{o}(k) = q_{\text{G}}(k)\\ 
 & \quad q_{o, \min} \le q_{o}(k) \le q_{o, \max}, \, o = 1, 2, \ldots, {|\cal O|}.
\end{aligned}
\end{equation}

The first constraint in \eqref{eq:OAP} holds by definition, while in the second constraint individual $q_o(k)$ input flows are subject to minimum and maximum permissible outflows. This is a static optimisation quadratic programming problem that can be efficiently solved by commercial or public available software. It should be noted that if the minimum and/or maximum bounds are activated then part of the global perimeter-ordered flow $q_{\text{G}}(k)$ will be wasted. On the other hand, if bound constraints are not activated then the optimisation yields the analytical solution:
\begin{equation}
\begin{bmatrix}\q(k)\\
 \lambda\end{bmatrix} = {\cal A}^{\dag}{\cal B}, \label{eq:solOpt}
\end{equation}
where $\q(k) \triangleq \u(k) \in \E^{|\cal O|}$ is the decision vector with elements the individual input flows $q_o(k)$, $o = 1, 2, \ldots, {|\cal O|}$; $\lambda$ is a Lagrange multiplier associated with the equality constraint in \eqref{eq:OAP}; ${\cal A}^{\dag}$ is the pseudoinverse matrix that arises in standard minimum norm approximation problems; and ${\cal B}$ is a vector with elements the right-hand side constants of \eqref{eq:OAP}. Precisely, ${\cal A}$ and ${\cal B}$ are given by:
\begin{align}
{\cal A} =  \begin{bmatrix}{\bf I}_{m\times m} & - \hat \q_{m\times 1}\\  {\bf 1}_{m\times 1}^\T & 0\end{bmatrix}, \quad\quad\quad
{\cal B} =  \begin{bmatrix}\hat{\q}_{m\times 1} \\ q_\text{G}(k)\end{bmatrix},
\end{align}
where $\hat{\q} \triangleq \hat{\u}$ is the vector of nominal input flows.

\section{Application, results, and discussion}\label{sec:CaseStudy}

\begin{figure*}[tbp]\centering
\begin{tabular}{cc}
\includegraphics[width=1.1\columnwidth]{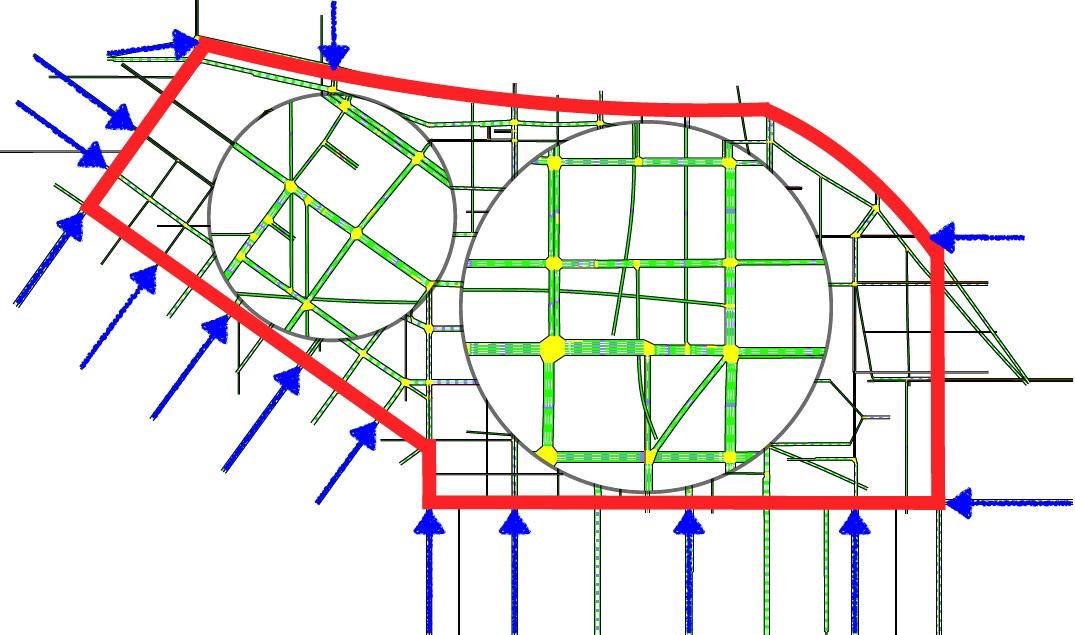} &
\includegraphics[width=0.9\columnwidth]{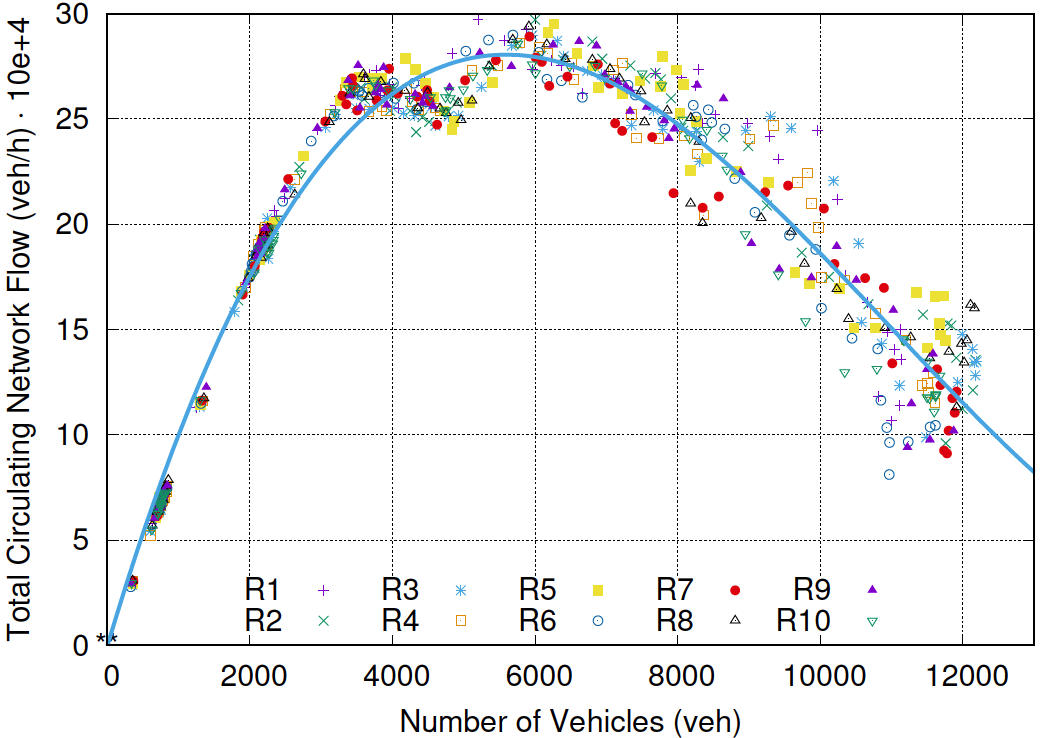}\\
\small (a) & \small (b)
\end{tabular}
\caption{Description of downtown San Francisco: (a) Protected network area (boundary in red colour) and fifteen controlled gates of entrance (illustrated by the blue arrows); (b) Network fundamental diagram of downtown San Francisco, adopted by \cite{KA_NG_PartB:2013}. R1-R10 denote ten different replications in AIMSUN microscopic simulator to reproduce the NFD.}\label{fig:SF_MFD}
\end{figure*}

\subsection{Network description}\label{sec:CaseSF}
Figure \ref{fig:SF_MFD}(b) depicts the shape of $O_c$ in function of vehicle accumulation $n(t)$ for a 2.5 square mile area of Downtown San Francisco, CA \citep{KA_NG_PartB:2013}, including about 110 junctions and 440 links with varying length from 160 m to 520 m (see Figure \ref{fig:SF_MFD}(a)). Fifteen entrance links and controlled gates are illustrated with blue arrows in  Figure \ref{fig:SF_MFD}(a).  Figure \ref{fig:SF_MFD}(b) confirms the existence of a fundamental diagram like-shape for the study area, which shape is seen to depend on the accumulation of vehicles. It can be seen that as the vehicle accumulation is increased from zero, the network flow increases to a maximum (flow capacity) and then turns down and decreases sharply to a low value possibly zero (in case of gridlock). Circulating flow capacity (around $27\times 10^4$ to $30\times 10^4$ veh/h) is observed at a varying vehicle accumulation in the range $[4000, 6000]$ veh. The shape of the fundamental diagram (and its critical parameters) was reproduced under different OD scenarios with dynamic traffic assignment  (C-Logit route choice model) activated every 3 min to capture somewhat adaptive drivers in a micro-simulation study via AIMSUN \citep{KA_NG_PartB:2013}. As can be seen, driver adaptation creates fundamental diagrams with less hysteresis that represent better real-life conditions (see also \citet{Mahmassani2013}). The shape of $O_c$ in Figure \ref{fig:SF_MFD}(b) can be approximated by the following 3rd order polynomial: 
\begin{equation}\label{eq:NFD}
O_c(n) =  4.128\times10^{-7}  n^3 - 0.0136 n^2 + 113.264 n,
\end{equation}
where $n \in \left[0, 13000\right]$ veh. To determine the output $O$ from $O_c$ via  $O[n(t)] \triangleq (l/L) O_c(n(t))$ an average trip length $L = 1.75$ km and average link length $l = 0.25$ km were considered. The utilised value of $L$ is consistent with the average trip length and the travel time across the protected network area of downtown San Francisco  \citep{KA_NG_PartB:2013}.

To account for potential excessive queues and corresponding waiting times outside of the protected network area, the fifteen gates shown in Figure \ref{fig:SF_MFD}(a) (illustrated with blue arrows) were included in the simulation and control models. Table \ref{tbl:GatesC} provides the different geometric characteristics of the fifteen entrance links and controlled gates. As can be seen, different gates have different length, vehicle storage capacities and other characteristics i.e., number of lanes, saturation flows. The consideration of gates with different geometric characteristics helps to have a meaningful comparison between the protected network and outside areas. Table \ref{tbl:GatesC} also provides the nominal $\hat{q}_{o}$, minimum $q_{o, \min}$ and maximum $q_{o, \max}$ permissible outflows for each $o \in {\cal O}$. These can easily be determined given saturation flows, minimum and maximum green times, and cycle times from a nominal traffic signal plan at each controlled gate of the protected network, as shown in Table \ref{tbl:GatesC}.

\begin{table*}[tbp]\centering
\caption{Different geometric characteristics of entrance links and controlled gates.}\label{tbl:GatesC}
\includegraphics[width=.9\textwidth]{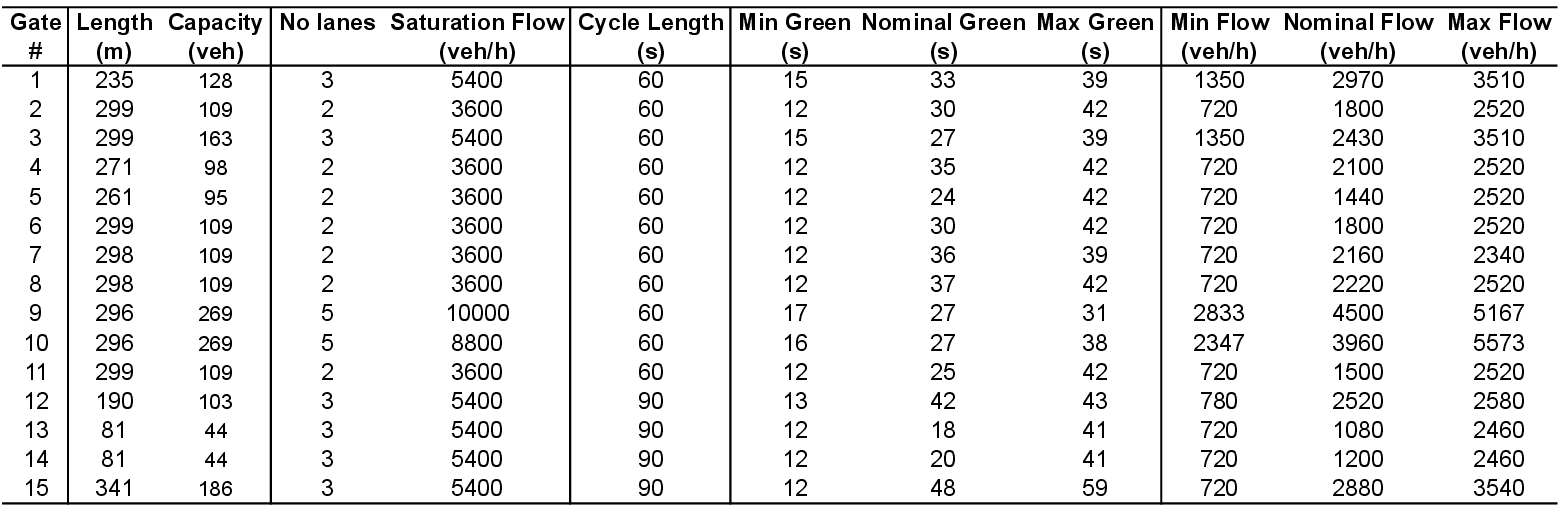}
\end{table*}

\subsection{Rolling horizon control design}\label{sec:cdesign}

Table \ref{tbl:GatesC} provides the different geometric characteristics of the fifteen ($|{\cal O}| = 15$) entrance links and controlled gates shown in Figure \ref{fig:SF_MFD}(a) (illustrated with blue arrows), thus $\x \in \mathbb{R}^{16}$ and $\u \in \mathbb{R}^{15}$, while $\hat{\ell}_o = 0$, $\forall\, o \in O$. The third column provides the storage capacity of each controlled link that is the vector $\x_{\max} \triangleq \begin{bmatrix}n_{\max} & \ell_{1,\max} & \cdots & \ell_{|{\cal O}|,\max}\end{bmatrix}^\T  \in \mathbb{R}^{16}$. The last three columns of the table provide the vectors $\u_{\min} = \q_{\min}$, $\hat{\u} = \hat{\q}$, and $\u_{\max} = \q_{\max}$, respectively. These values are calculated from the field applied signal plans presented in columns five ($S_o$: saturation flow), six ($C$: cycle length) , seven ($g_{o,\min}$: minimum green time), eight ($\hat{g}_{o}$: nominal green time), and nine ($g_{o,\max}$: maximum green time), via $gS/C$. In this way, any input flows ordered by the multi-gated perimeter flow control strategy are feasible traffic signal plans.  Note that traffic signals at controlled gates are all multistage fixed-time operating on a common cycle length of 90 s for the west boundary of the area (The Embarcadero including gates $o = 1, 2, \ldots, 11$) and 60 s for the rest of the network (gates $o = 12, 13, \ldots, 15$). 

For the solution of \eqref{eq:QP_without_C} or \eqref{eq:QP}--\eqref{eq:QP_with_C} it suffices to specify the state matrices $\A \in \mathbb{R}^{16\times 16}$, $\B\in \mathbb{R}^{16\times 15}$, and $\C\in \mathbb{R}^{16\times 16}$, and weighting matrices $\Q\in\mathbb{R}^{16\times 16}$ and $\R$. All state matrices are constructed for the studied network on the basis of the selected $\hat{\x}^\T = [\hat{n}\quad \0] \in \mathbb{R}^{16}$, $\hat{\u} =\hat{\q} \in \mathbb{R}^{15}$ and $\hat{\d} = \0$, and sampling time $T = 180$ s. More precisely, $\A = \diag(1- 4.94 \times T, 1, \ldots, 1)$, $\B = T \begin{bmatrix}{\bf 1}_{1\times 15} & -{\bf I}_{15\times 15}\end{bmatrix}^\T$, and $\C = \diag(T, \ldots, T)$. The matrix $\Q = \diag(1/w, 1/\ell_{1,\max}, \ldots, 1/\ell_{|{\cal O}|,\max})$ is selected, where $w = 2000$ veh was found appropriate to achieve equity (see Section \ref{sec:multigated} for a conceptual explanation). The diagonal elements of $\R$ were set equal to $r = 0.00001$ after a trial and error procedure. The disturbance vector $\d$ consists of the demands $d_o$, $o = 1,\ldots, 15$,  at every origin of the protected network and disturbance $d_n$ of the fundamental diagram.

\subsection{Sensitivity analysis on the NFD parameters and optimisation horizon $N_o$}

\begin{figure*}[tbp]
\centering
\begin{tabular}{ccc}
\includegraphics[width=0.6\columnwidth]{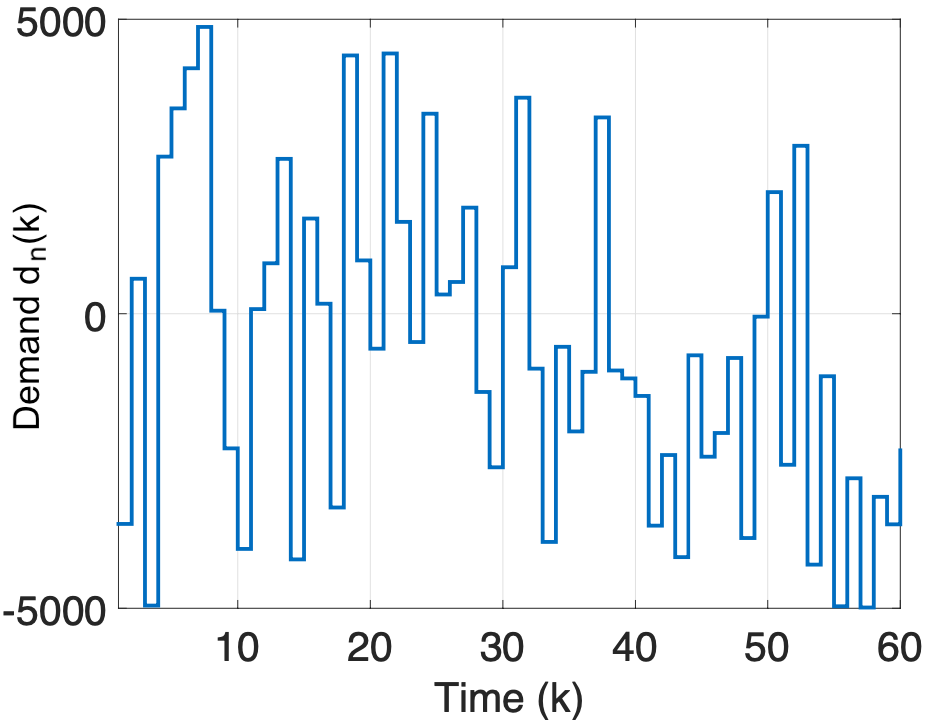} &
\includegraphics[width=0.69\columnwidth]{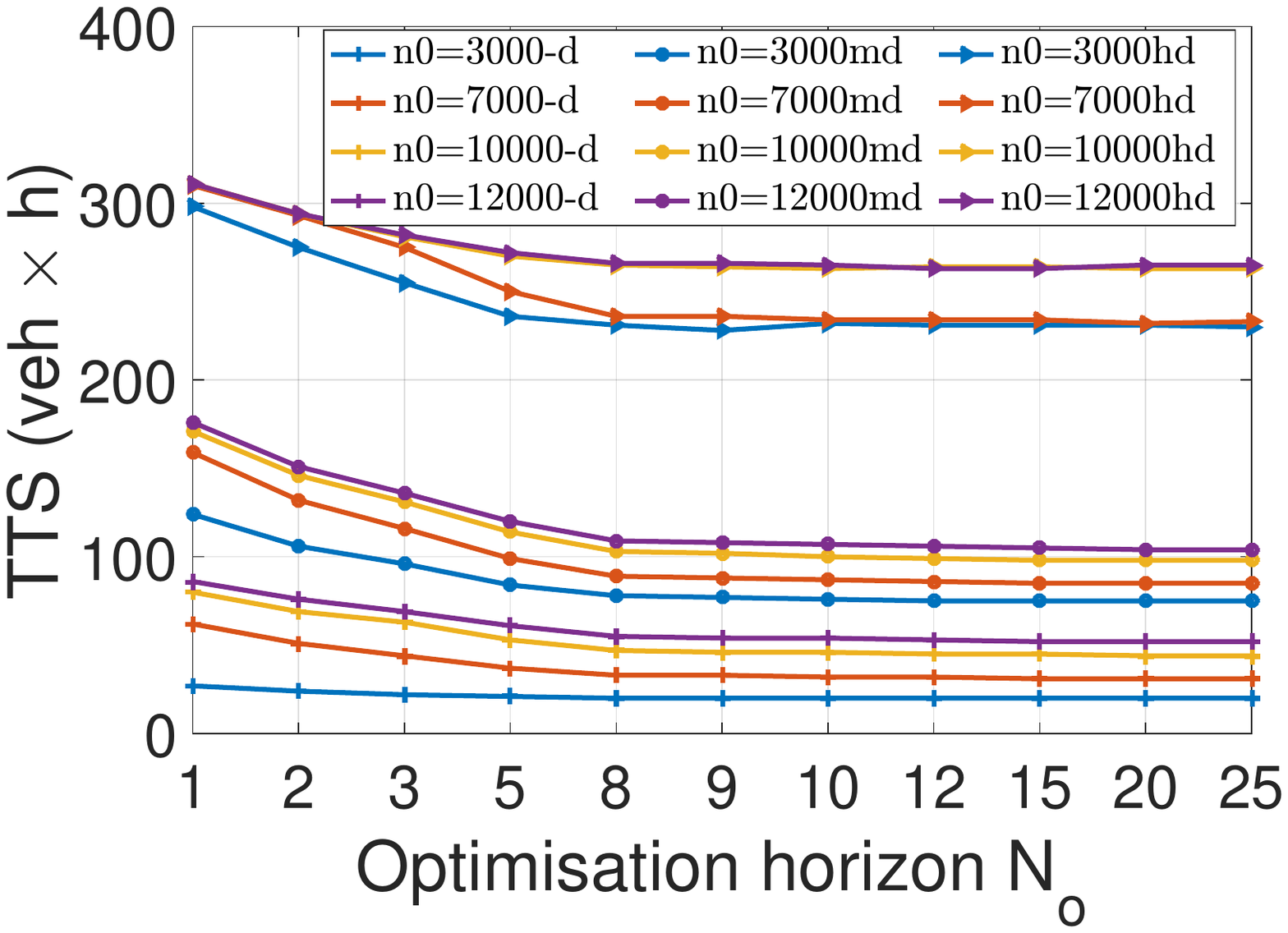} & 
\includegraphics[width=0.69\columnwidth]{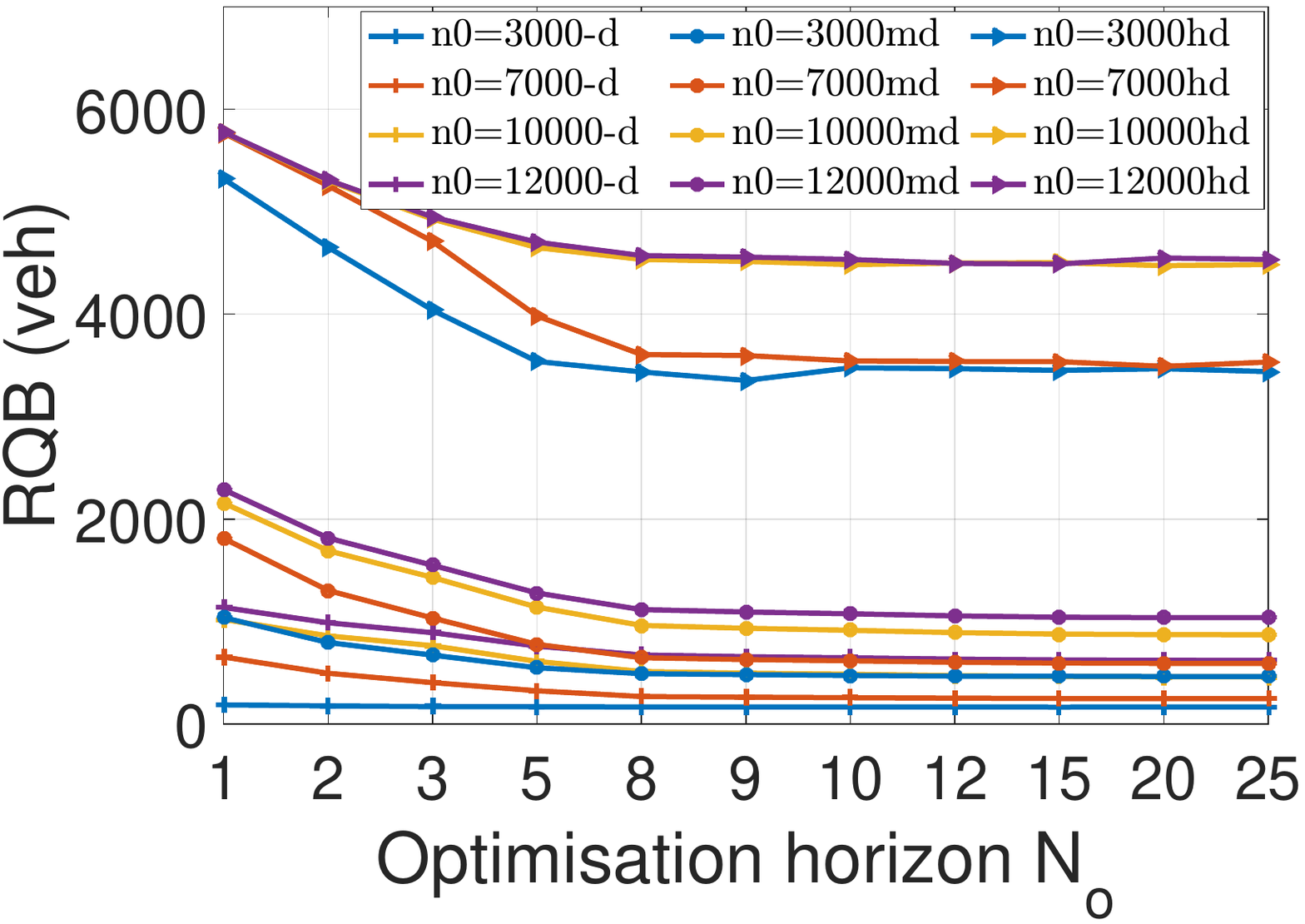}\\
 \small (a)  & \small (b) & \small (c)
\end{tabular}
\caption{(a) Demand/disturbances within the protected network area;  (b) and (c) Sensitivity analysis and performance of multi-gated perimeter flow control for different optimisation horizon $N_o$ and different demand scenarios: (b) TTS within the protected network and outside network area; (c) RQB within the protected network and outside network area; Each line in the legend of the figures indicates a different initial state $n(0)$ veh for a different demand scenario (-d: no demand, md: medium demand, hd: high demand).}\label{fig:SelectionNo}
\end{figure*}

The desired vehicle accumulation for \eqref{eq:linear_v_dt} is selected $\hat{n} = 4000$ veh after sensitivity analysis for $\hat{n} = \{4000, 4500, 5000\}$ veh, while $c=0.9$ is set in \eqref{eq:Realoutflow} after sensitivity analysis for $c = \{0.85, 0.90, 0.95\}$. The selected parameters values found appropriate to stabilise the NFD around its desired point $\hat n$ and prevent overflow phenomena within the protected network area (see eq.\ \eqref{eq:Realoutflow}).  Furthermore, twelve scenarios were defined in order to investigate the behaviour of the multi-gated perimeter flow control strategy under different initial states and demand scenarios. The twelve scenarios composed of four initial states in the uncongested and congested regimes (near gridlock traffic conditions) of the fundamental diagram $n(0) = \{3000, 7000, 10000, 12000\}$ veh and three different demand scenarios namely no external demand ($\d = \0$), medium demand, and high demand. For the medium and high scenarios trapezoidal demands have been used for $d_o(k)$, $o = 1,\ldots, 15$, $k = 0, \ldots, N_p-1$ over the predicted horizon of $N_o = N_p$. To capture the uncertainty of the (scaled) fundamental diagram, particularly when the network is operating in the congested regime (notice the noise for $n>6000$ veh), $d_n$ is selected to vary randomly with respect to $n(k)$ in the range $[-5000, 5000]$ veh/h for $n > 6000$ veh (see Figure \ref{fig:SelectionNo}(a)). The rolling-horizon strategy is run with different optimisation horizons $N_o = \{1, 2, 3, 5, 8, 9, 10, 12, 15, 20, 25\}$ in order to investigate the impact of $N_o$ on the control performance. 

For each of the twelve scenarios and for each demand scenario, two evaluation criteria are calculated for comparison. The total time spent
\begin{equation}
  {\rm TTS} \triangleq T \sum_{k=0}^{N_o} \left(\sum_{o =1}^{|\cal O|} \ell_o(k) + n(k)\right) \qquad \mbox{(in veh}\times\mbox{h)}
\end{equation}
and the relative queue balance
\begin{equation}
  {\rm RQB} \triangleq \sum_{k=0}^{N_o} \left( \sum_{o =1}^{|\cal O|}
  \frac{{\ell_o(k)}^2}{\ell_{o,\max}} +  \frac{{n(k)}^2}{n_{\max}}\right) \quad\qquad \mbox{(in veh)}.
\end{equation}
within the protected network and outside network area, where $N_o$ is the scenario and optimisation time horizon.

Figures \ref{fig:SelectionNo}(b)--\ref{fig:SelectionNo}(c) display the obtained TTS and RQB results for the rolling-horizon MGC approach for different optimisation horizons $N_o = \{1, 2, 3, 5, 8, 9, 10, 12, 15, 20, 25\}$. Each line in the legend of the figures indicates a different initial state $n(0) \in \{3000, 7000, 10000, 12000\}$ veh under three different demand scenarios (-d: no demand, md: medium demand, and hd: high demand). It can be seen that for $N_o \ge 10$ there are no significant deviations of the evaluation criteria for different optimisation horizons $N_o$ even for the high-demand scenarios. The assessment criteria at the gated links are seen to improve as  $N_{o}$ increases in some scenarios. In particular, for the most congested accumulation ($n(0) =12000$ veh) and with high demand, the most satisfactory results with respect to both evaluation criteria are obtained with $N_{o} = 15$ (equivalent to 0.75 h = 45 minutes). Hence $N_{o} = 15$ equivalent to 45 minutes is a reasonable choice. In principle, a satisfactory optimisation horizon is the one that is in the order of the time needed to travel through the network (see e.g., \cite{Aboudolas2010}). 

It can be seen that for $N_o \ge 10$ there are no significant deviations of the evaluation criteria for different optimisation horizons $N_o$ even for the high-demand scenarios. The assessment criteria at the gated links are seen to improve as  $N_{o}$ increases in some scenarios. In particular, for the most congested accumulation ($n(0) =12000$ veh) and with high demand, the most satisfactory results with respect to both evaluation criteria are obtained with $N_{o} = 15$ (equivalent to 0.75 h = 45 minutes). Hence $N_{o} = 15$ equivalent to 45 minutes is a reasonable choice. In principle, a satisfactory optimisation horizon is the one that is in the order of the time needed to travel through the network (see e.g., \cite{Aboudolas2010}).

\subsection{Control results}\label{sec:OLresults} 

Several tests were conducted in order to investigate the behaviour of the proposed multi-gated control for different scenarios. The scenarios were created by assuming more or less high initial queues $\ell_o(0)$ in the fifteen origin links of the protected network while the protected network area operating in the congested regime, i.e. its state $n(0) > 6000$ veh. The optimisation horizon for each scenario is 2 h (40 cycles). Extensive simulation results for various demand scenarios can be found in \cite{Ruzanna:19}.

The calculated optimal state and control trajectories demonstrate the efficiency of the proposed multi-gated control to solve the perimeter flow control problem with queue dynamics. Figure\ \ref{fig:results} shows some obtained trajectories for a heavy scenario with $\ell_o(0) = 0.7 \ell_{o,\max}, \, \forall o =1, 2, \ldots, 15$ and two initial states in the congested regime of the fundamental diagram $n(0) = 7000$ veh and $n(0) = 12000$ veh (near gridlock traffic conditions). Tests were conducted with and without external demand flows at origin links, denoted with ``s+d" and ``s-d", respectively. The demand profile follows a trapezoidal pattern, starting with low values that gradually increase to high flows, matching the saturation flow at each origin link. These flows, constituting approximately 25\% to 40\% of the saturation flow at each origin link, are maintained for a specified duration before gradually decreasing back to low values. It should be highlighted that despite the same level of saturation (initial queues at 70\% of $\ell_{o,\max}$) being used for all origin links, the corresponding vehicle queues observed vary due to different geometric characteristics. The main observations are summarised in the following remarks:
\begin{itemize}
	\item  The MGC strategy manages to stabilise the vehicle accumulation of the protected network around its desired point $\hat{n} = 4000$ veh for all initial points (even in the extreme case) and cases with and without disturbances (see Figure \ref{fig:results}(c)). 
	\item The MGC manages to dissolve the initial origin link queues in a balanced way (see Figures \ref{fig:results}(g--o)) and thus, the desired control objective of queue balancing and equity for drivers using different gates to enter the protected network area is achieved.
	\item  The MGC strategy manages to stabilise all input flows to their desired values $\hat{\q}$ (corresponding to the nominal signal plan in Table \ref{tbl:GatesC}) in the steady state, i.e., where $n = \hat{n} = 4000$ and system's throughput is maximised (see Figures \ref{fig:results}(d--i), notice the different reference points $\hat{q}_o$ in each subfigure).   
	\item The input flows ordered by the multi-gated perimeter flow control strategy have different trajectories and characteristics (see control trajectories in Figures \ref{fig:results}(d--i)). This confirms that an equal distribution of ordered flows to corresponding junctions is not optimal, as largely assumed in previous studies. As can be seen, the proposed strategy determines optimally distributed input flows (or feasible entrance link green times) by taking into account the individual geometric characteristics of the origin links as well as minimum and maximum constraints.
	\item It is evident that excessive demand and high initial queues at origin links, coupled with the applied control, causes congestion shortly after the beginning of the time horizon. At the same time the protected network is operating in the congested regime ($n(0) = 7000$ veh or $n(0) = 12000$ veh). As can be seen, the multi-gated control strategy first restricts the high initial queues at origin links to flow into the oversaturated protected network area and then, in order to manage the developed long queues therein (in some cases reach the upper bounds), it gradually increases the input flows. Note that for some gates (7, 8 and 9) bound constraints are activated for a certain time period.             
\end{itemize}
\subsection{Comparison of MGC with single-region control}\label{sec:OLresults}

\begin{figure*}\centering
\begin{tabular}{ccc}
\includegraphics[width=55mm]{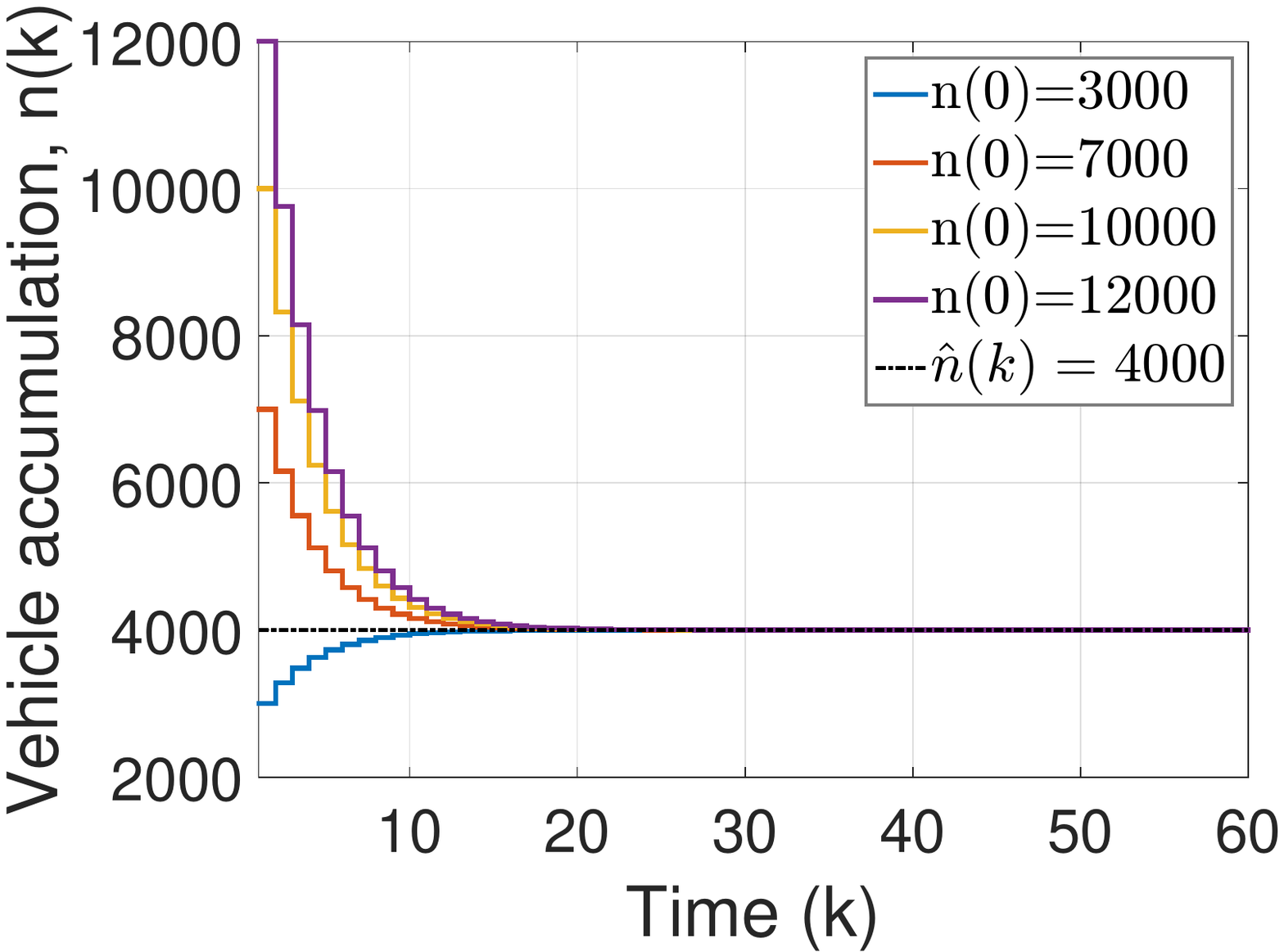} &    
\includegraphics[width=55mm]{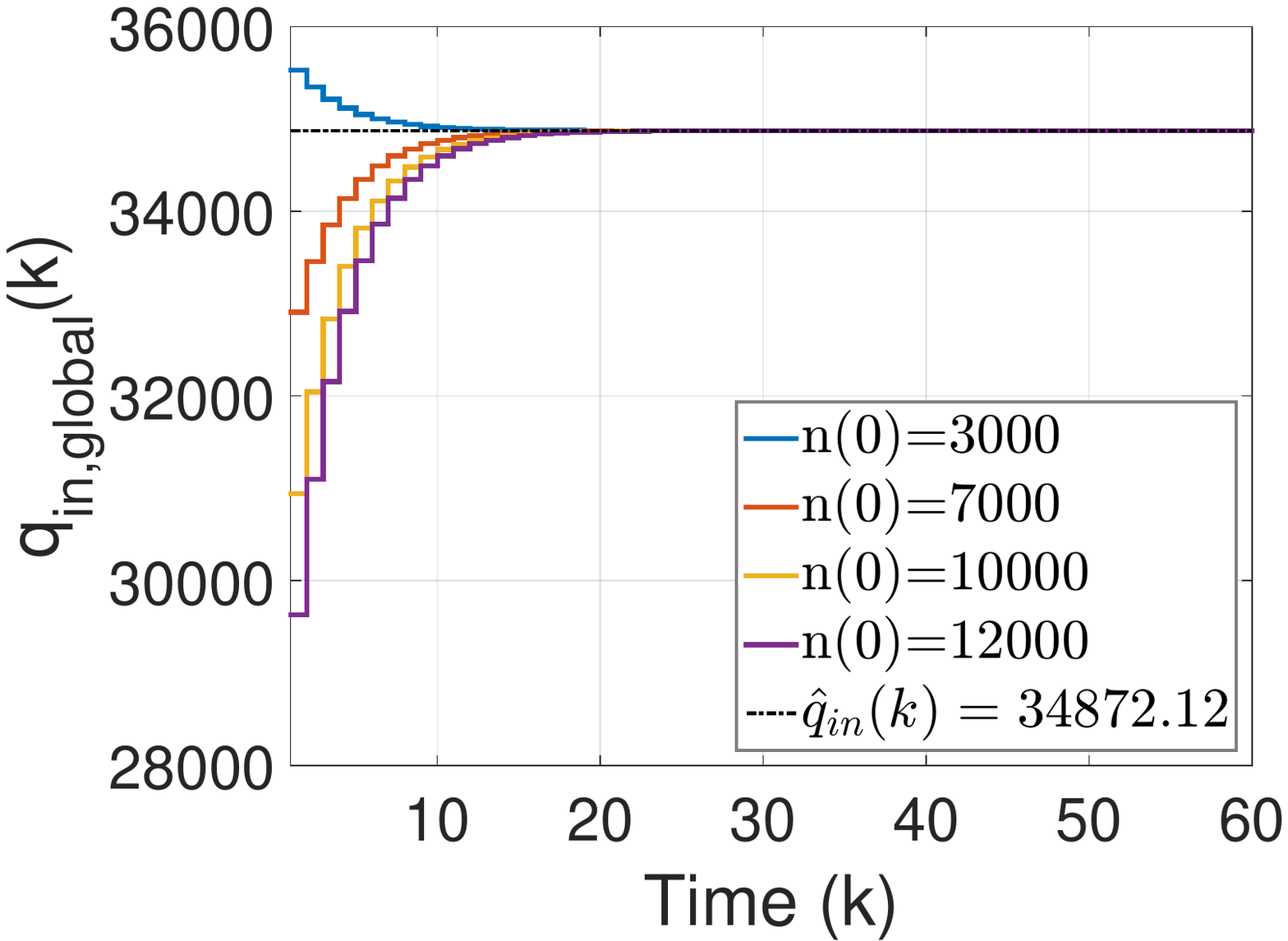}   & 
\includegraphics[width=55mm]{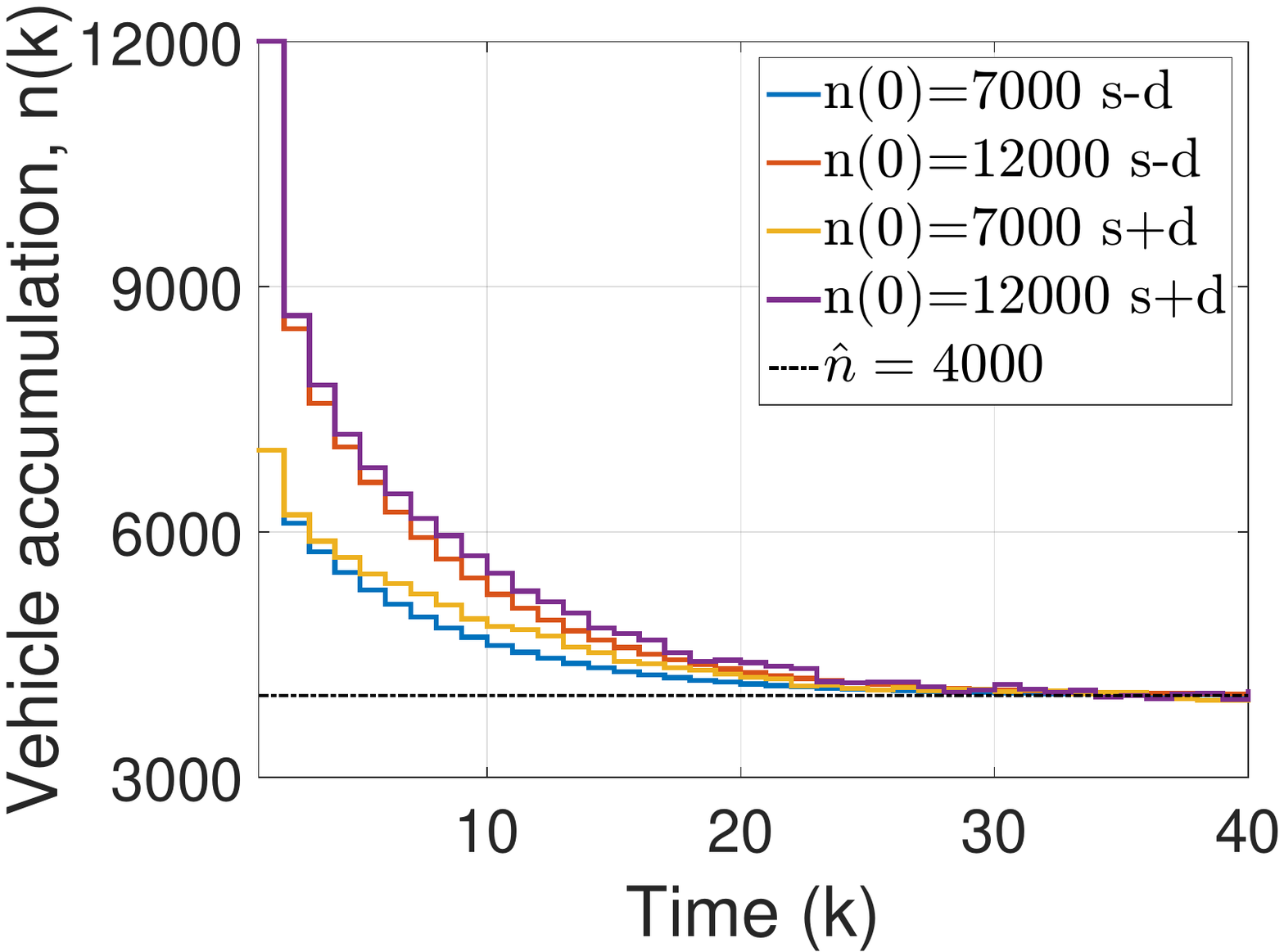} \\
(a) State trajectories (global input) & (b) Global input flow  &  (c) State trajectories (Multi-gatedl)  \\[2pt]
\includegraphics[width=55mm]{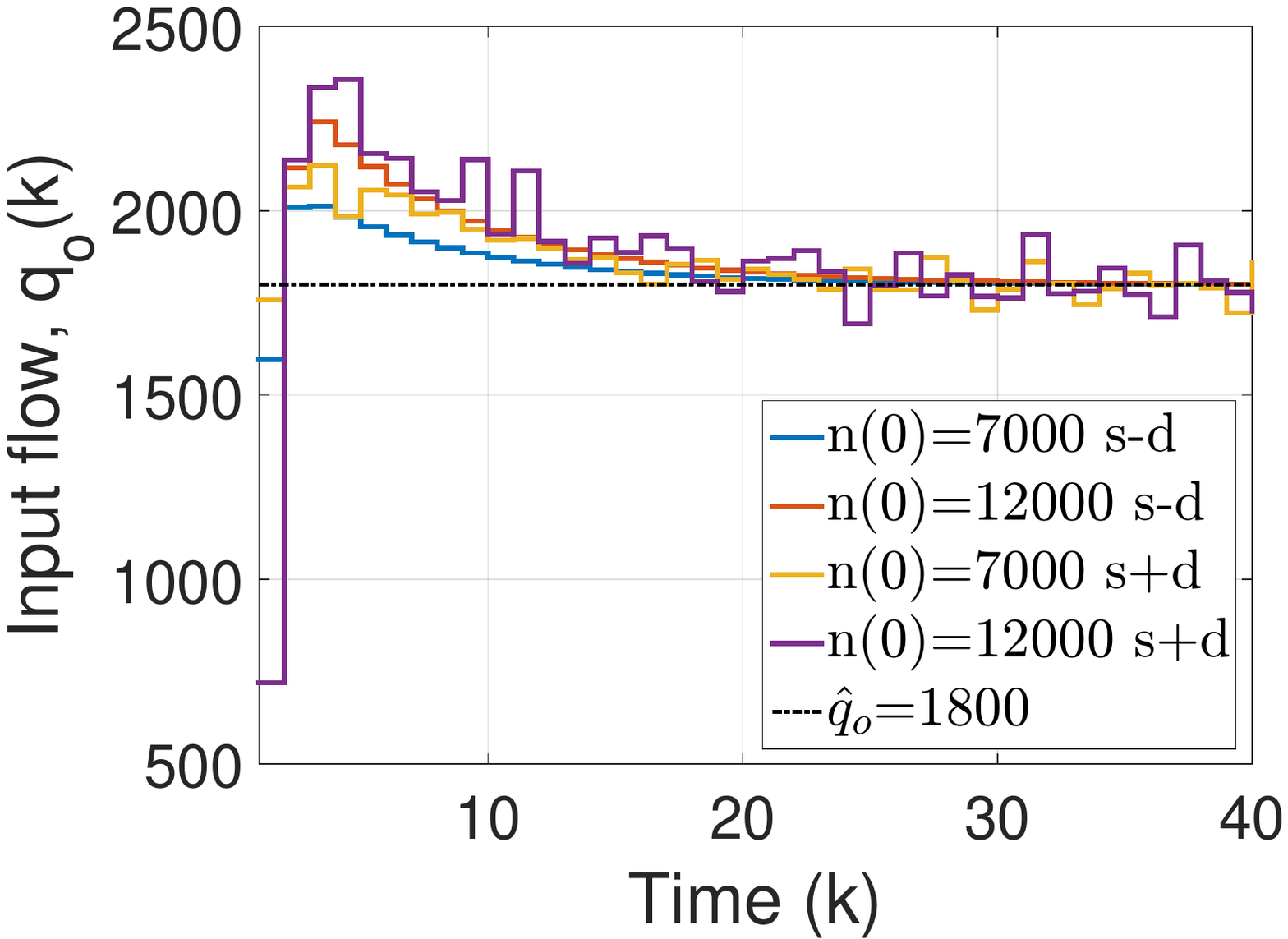}     &    
\includegraphics[width=55mm]{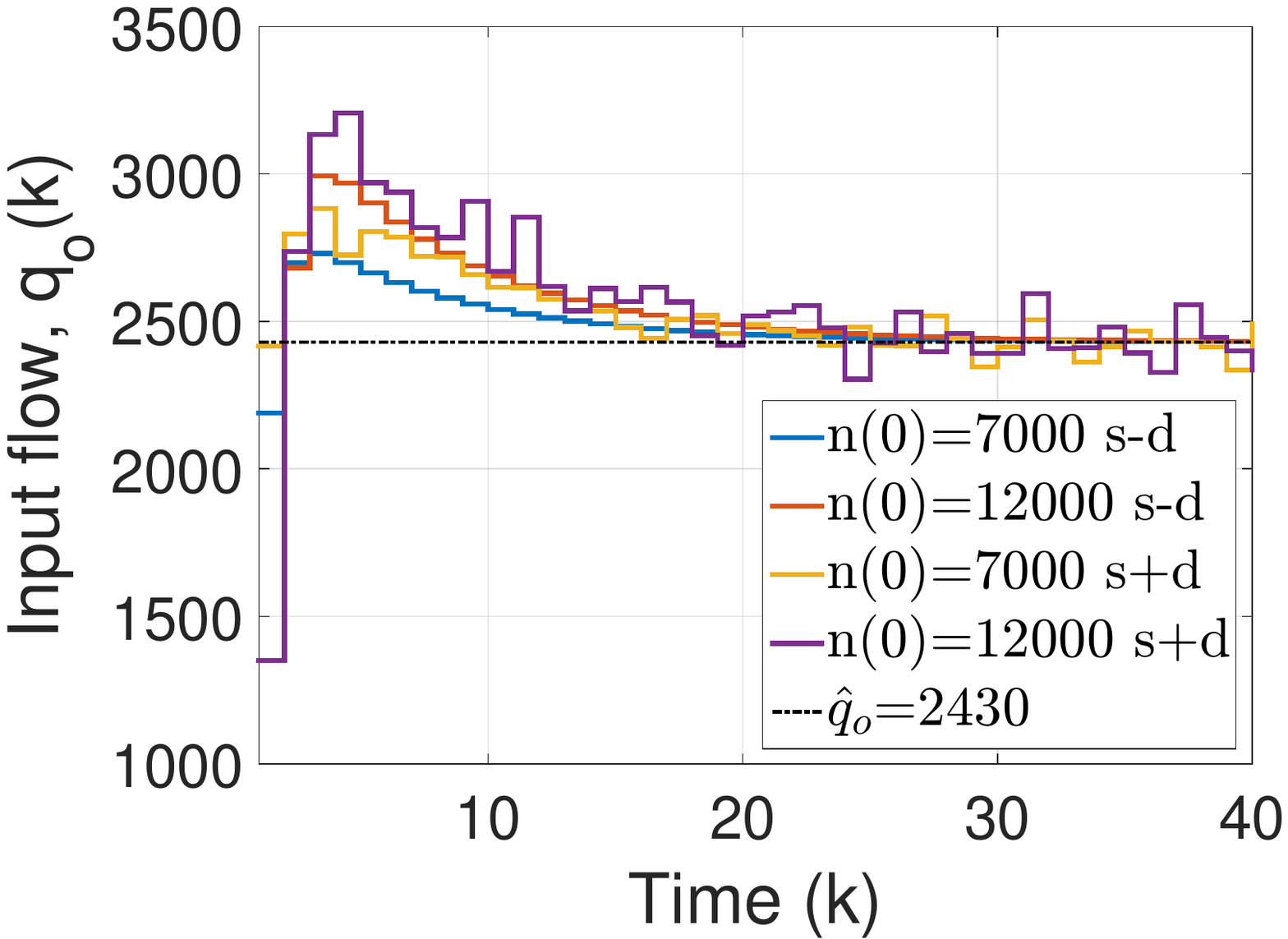}    &    
\includegraphics[width=55mm]{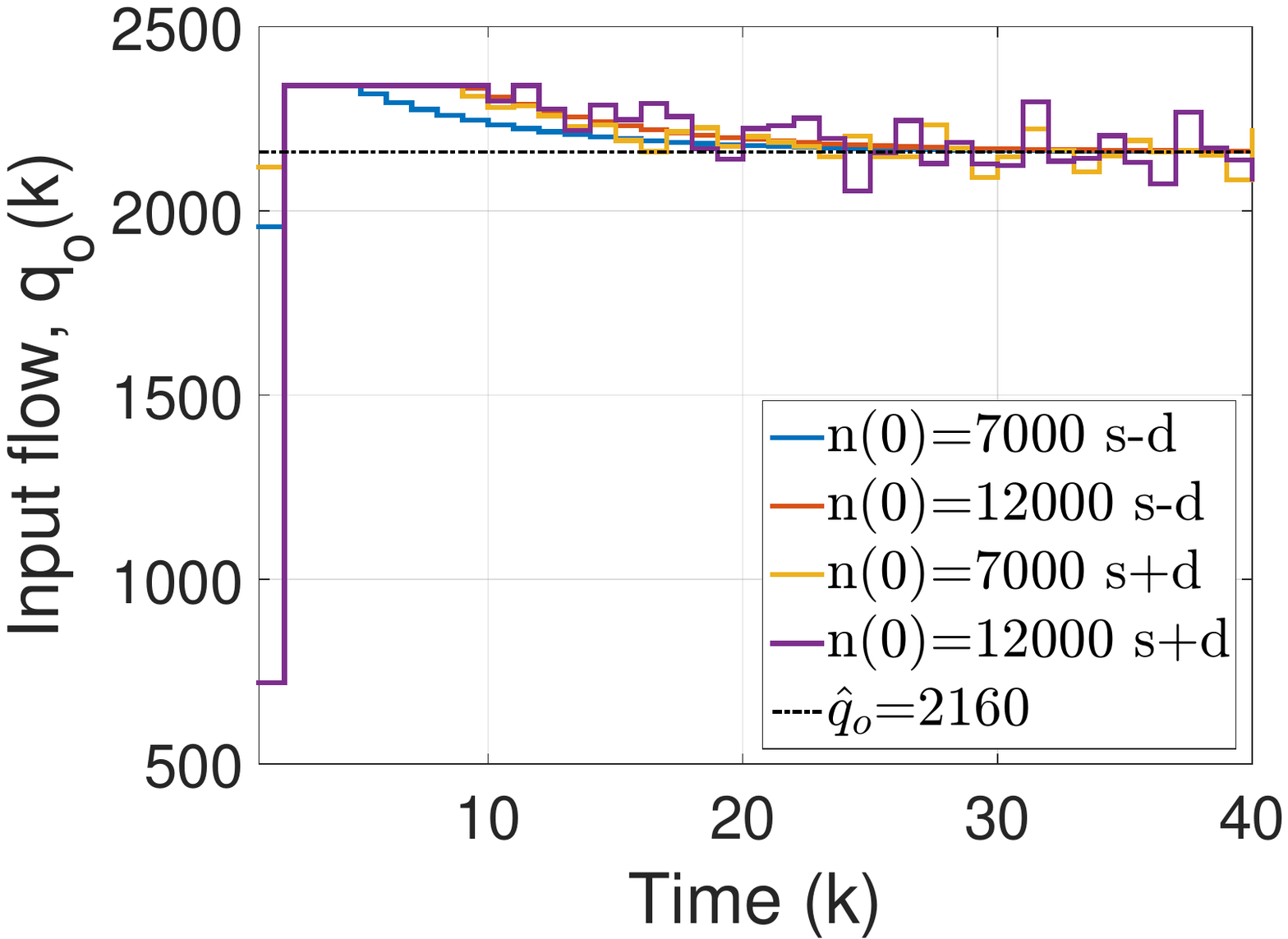} \\
(d) Input flow: Gate 2  &    (e) Input flow: Gate 3  &    (f) Input flow: Gate 7\\[2pt]
\includegraphics[width=55mm]{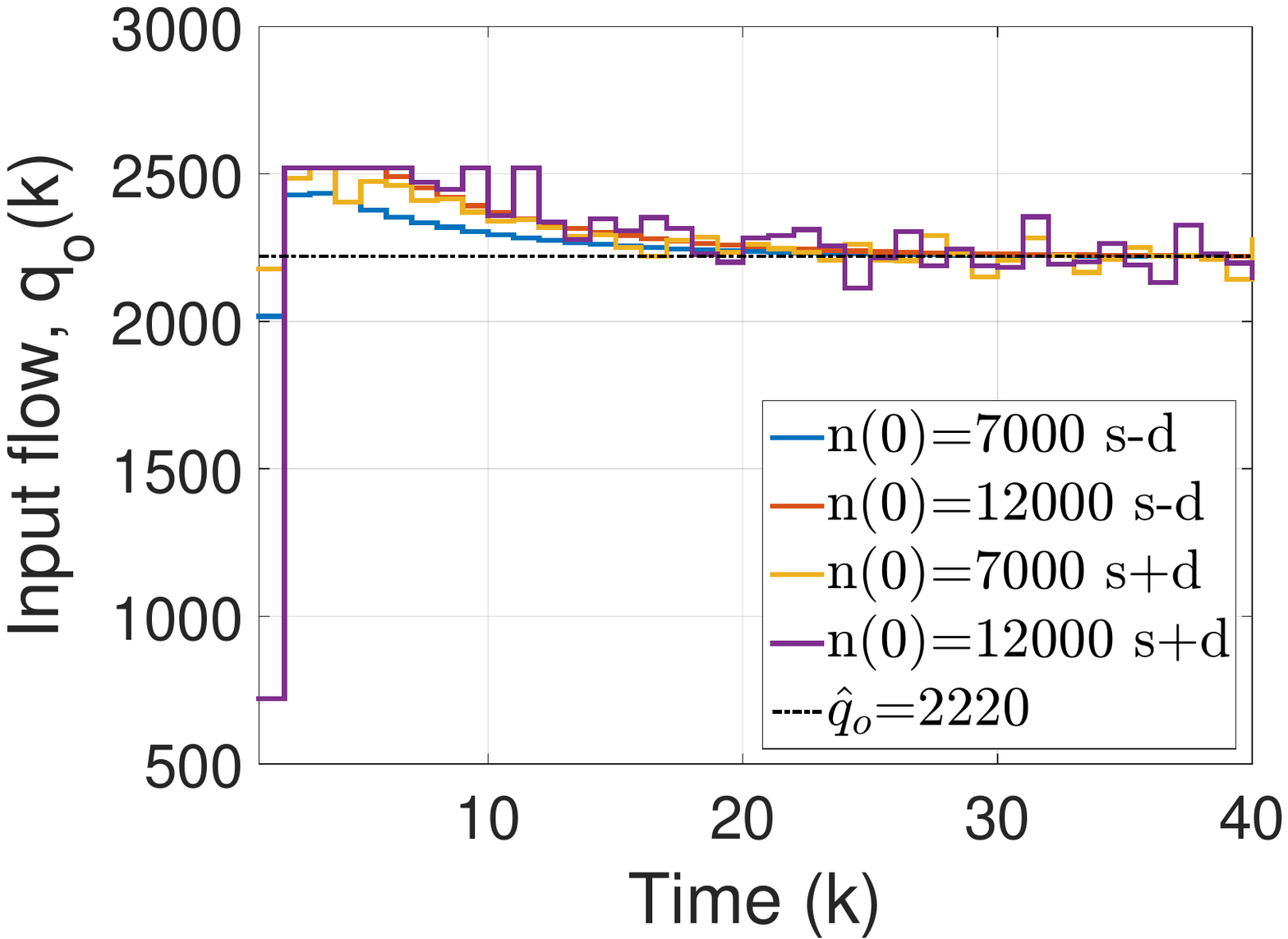}     &   
 \includegraphics[width=55mm]{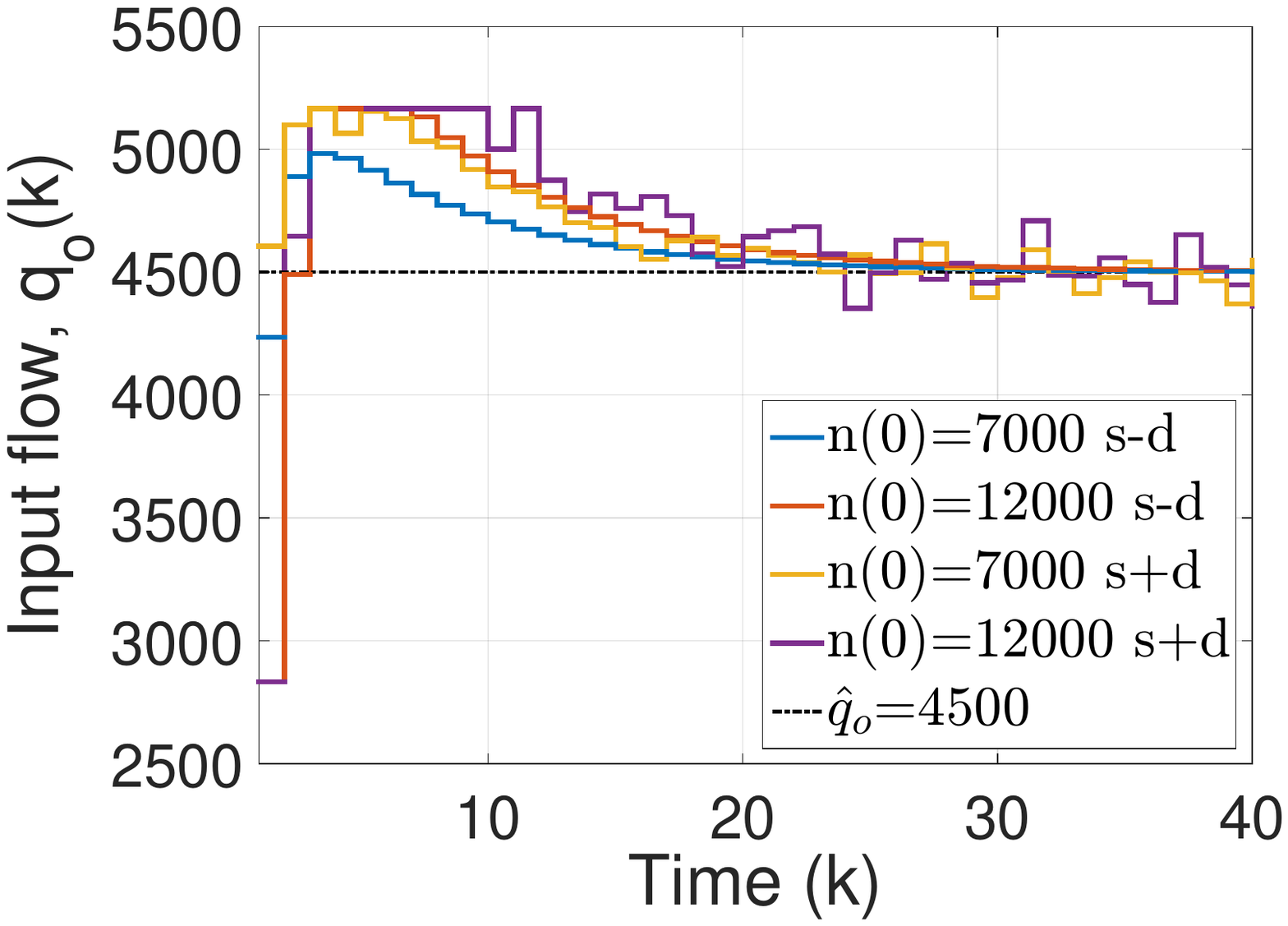}     &    
 \includegraphics[width=55mm]{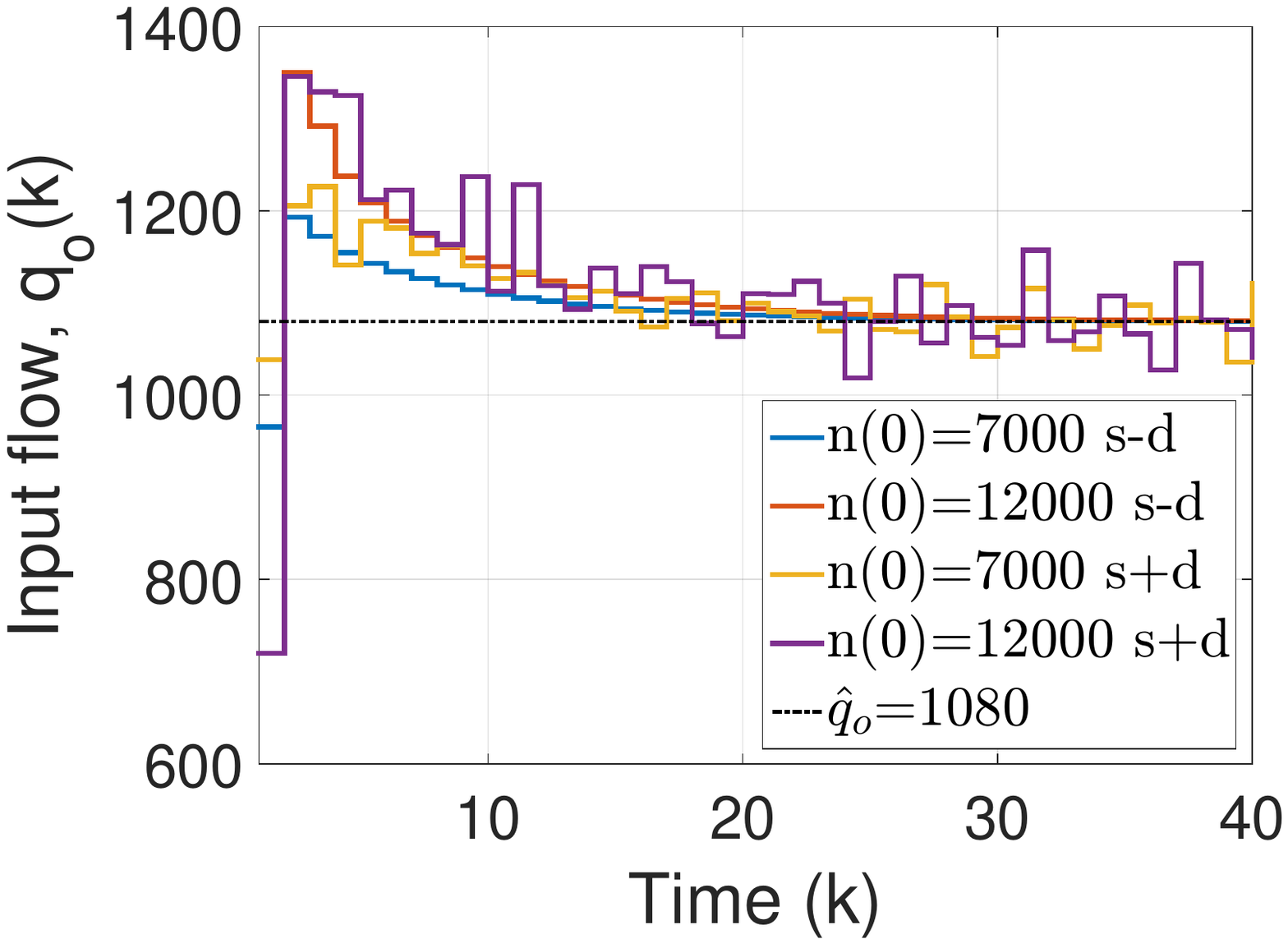} \\
(g) Input flow: Gate 8 &     (h) Input flow: Gate 9 &     (i) Input flow: Gate 13 \\[2pt]
\includegraphics[width=55mm]{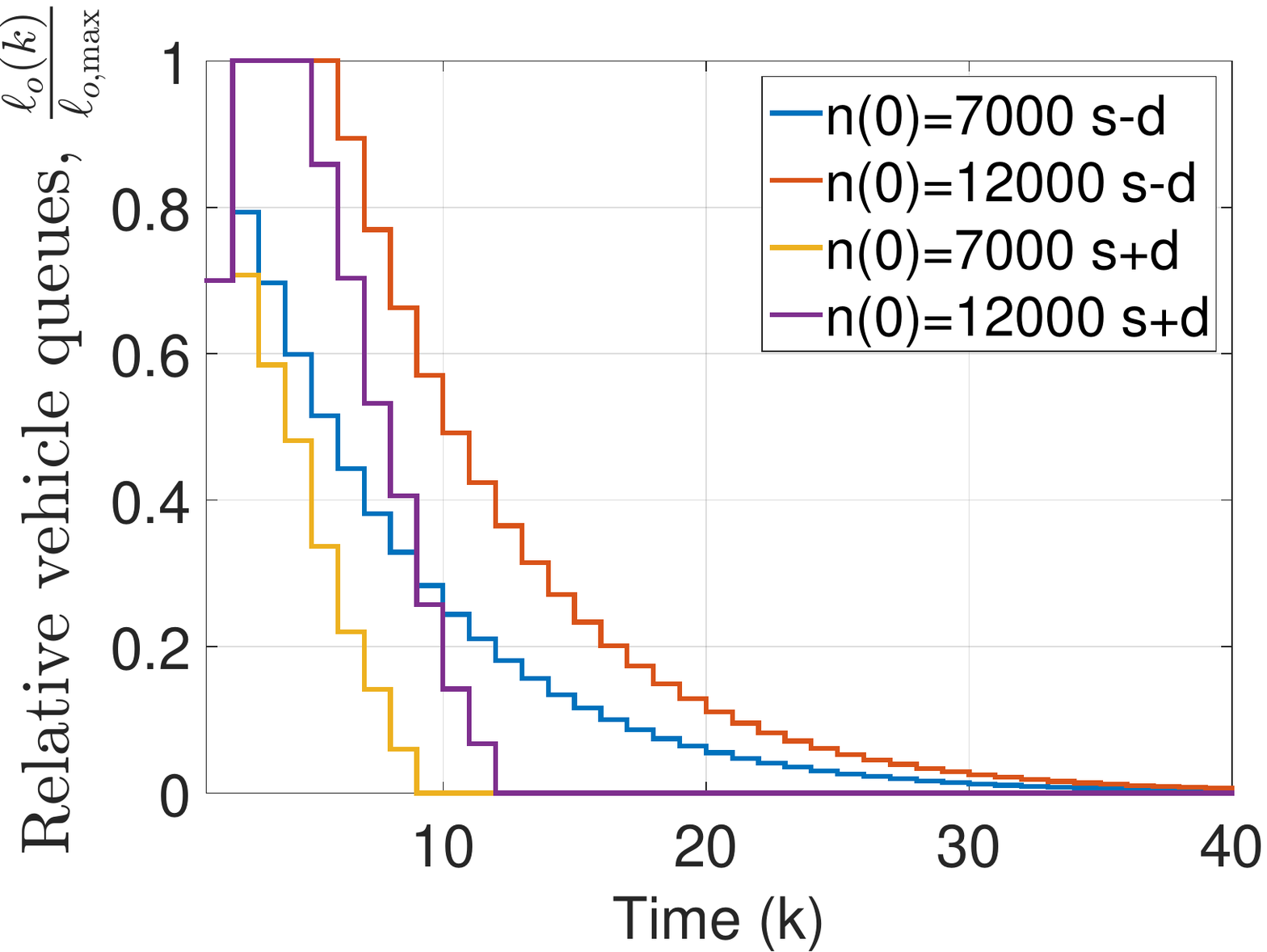}  &        
\includegraphics[width=55mm]{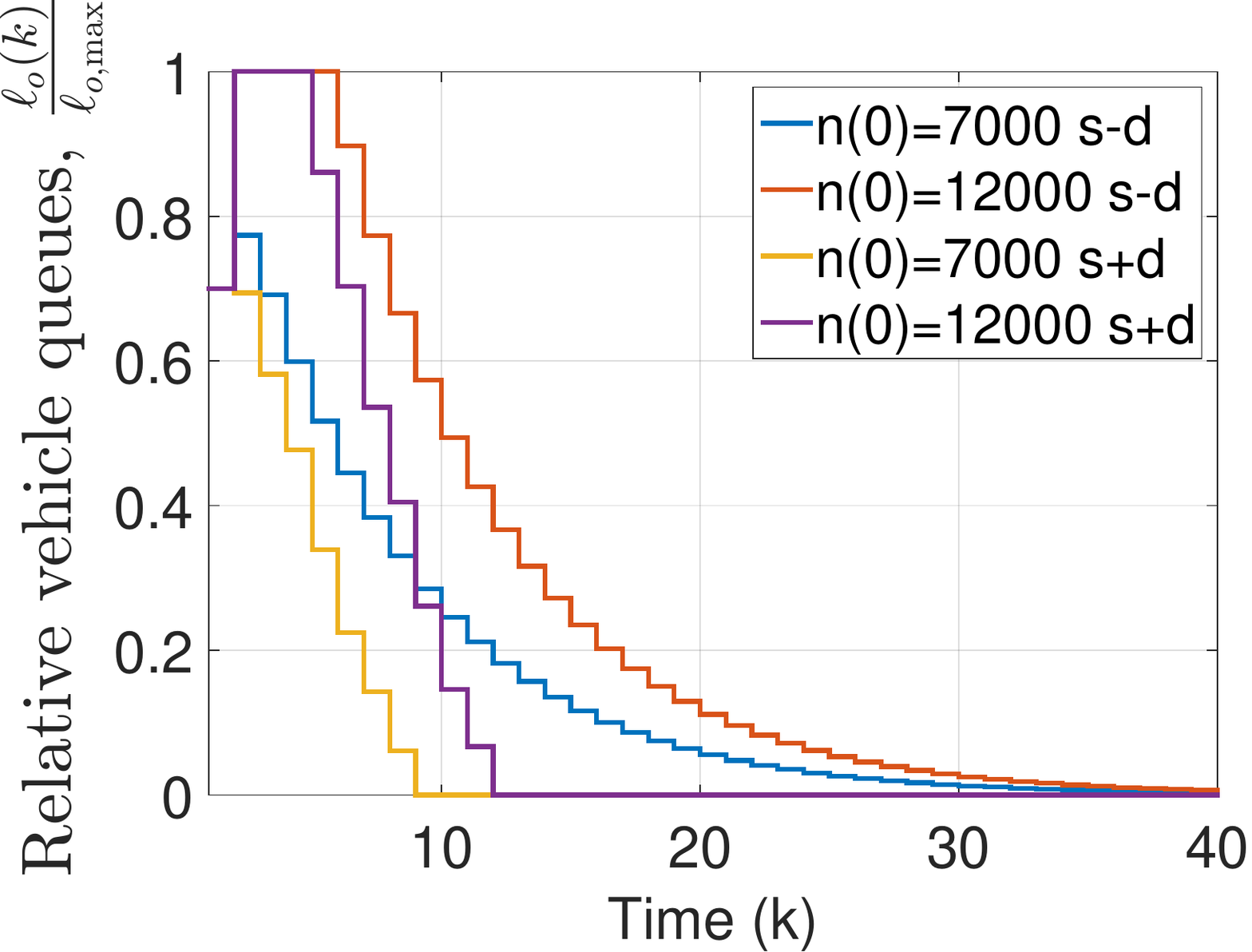}     &    
\includegraphics[width=55mm]{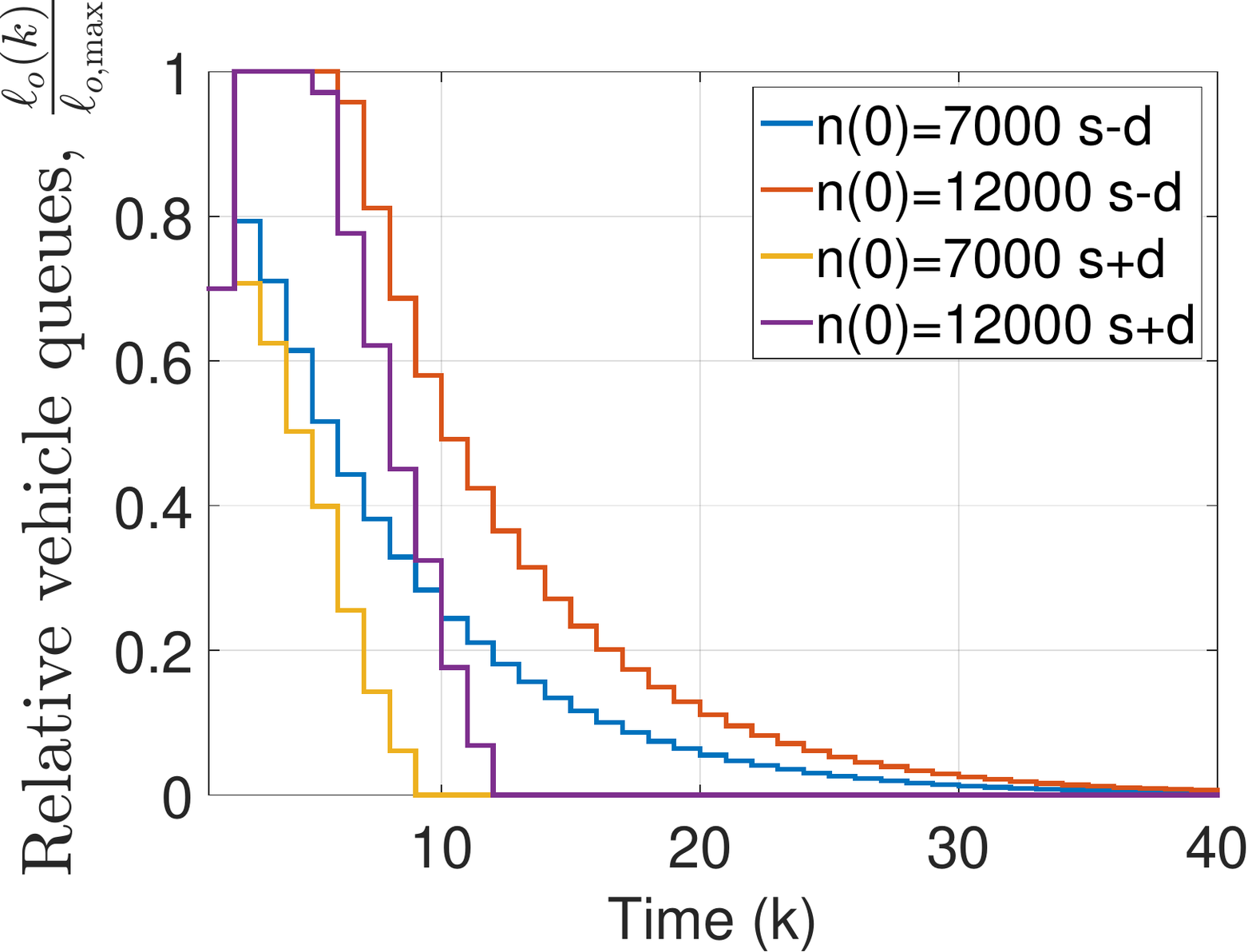} \\
(j) Relative queue: link 2  &    (k) Relative queue: link 3 &     (l) Relative queue: link 7 \\[2pt]
\includegraphics[width=55mm]{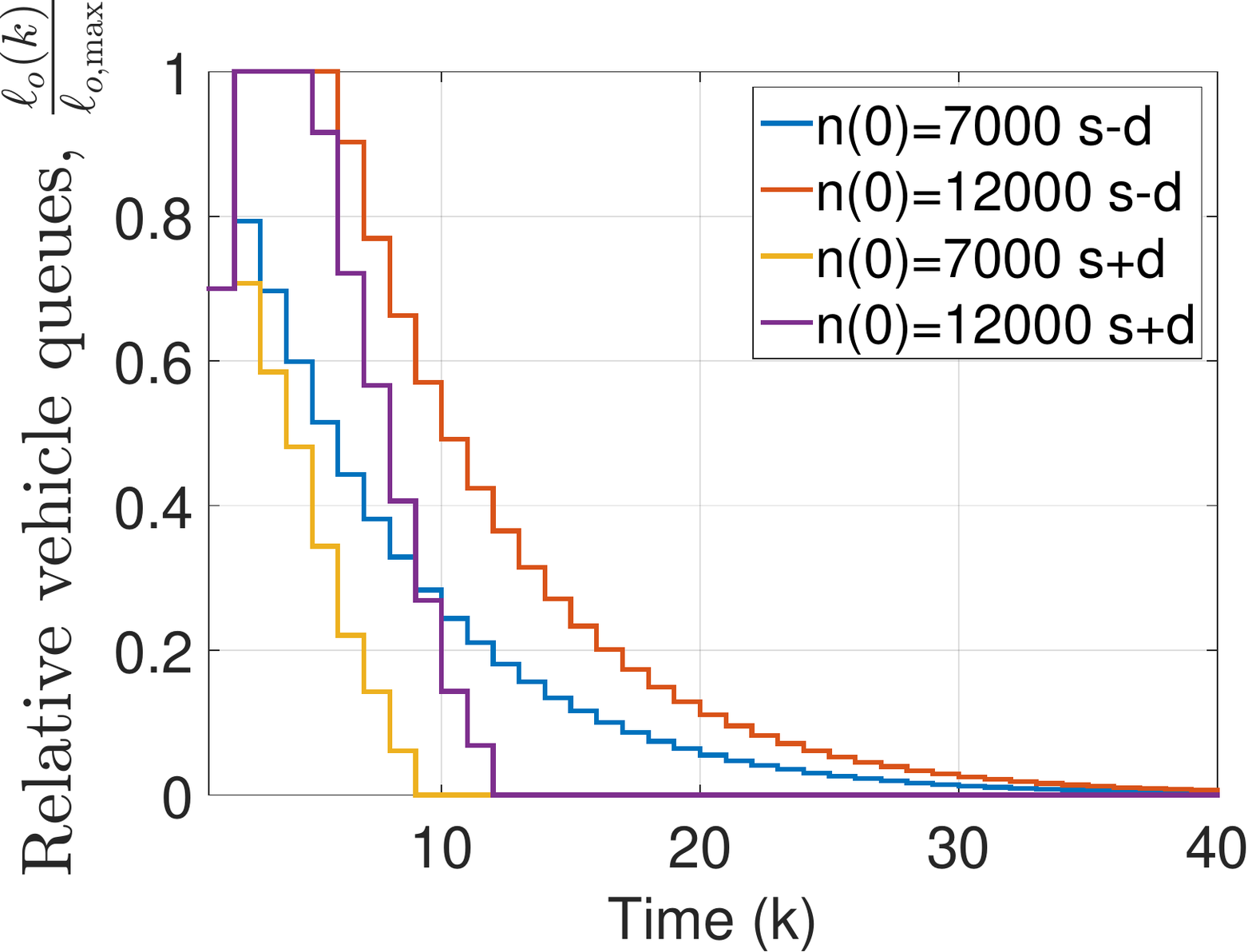}     &    
\includegraphics[width=55mm]{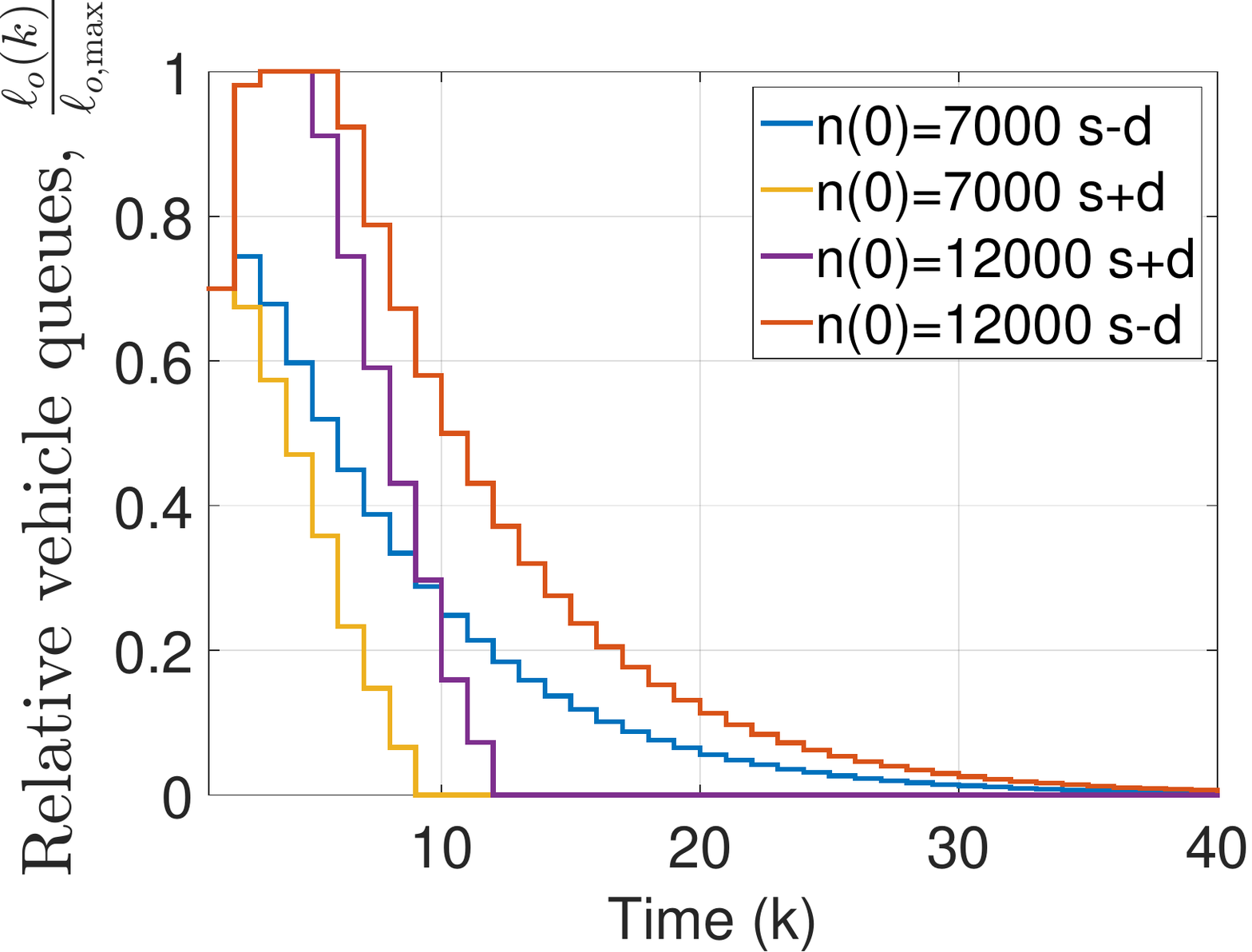}   &    
\includegraphics[width=55mm]{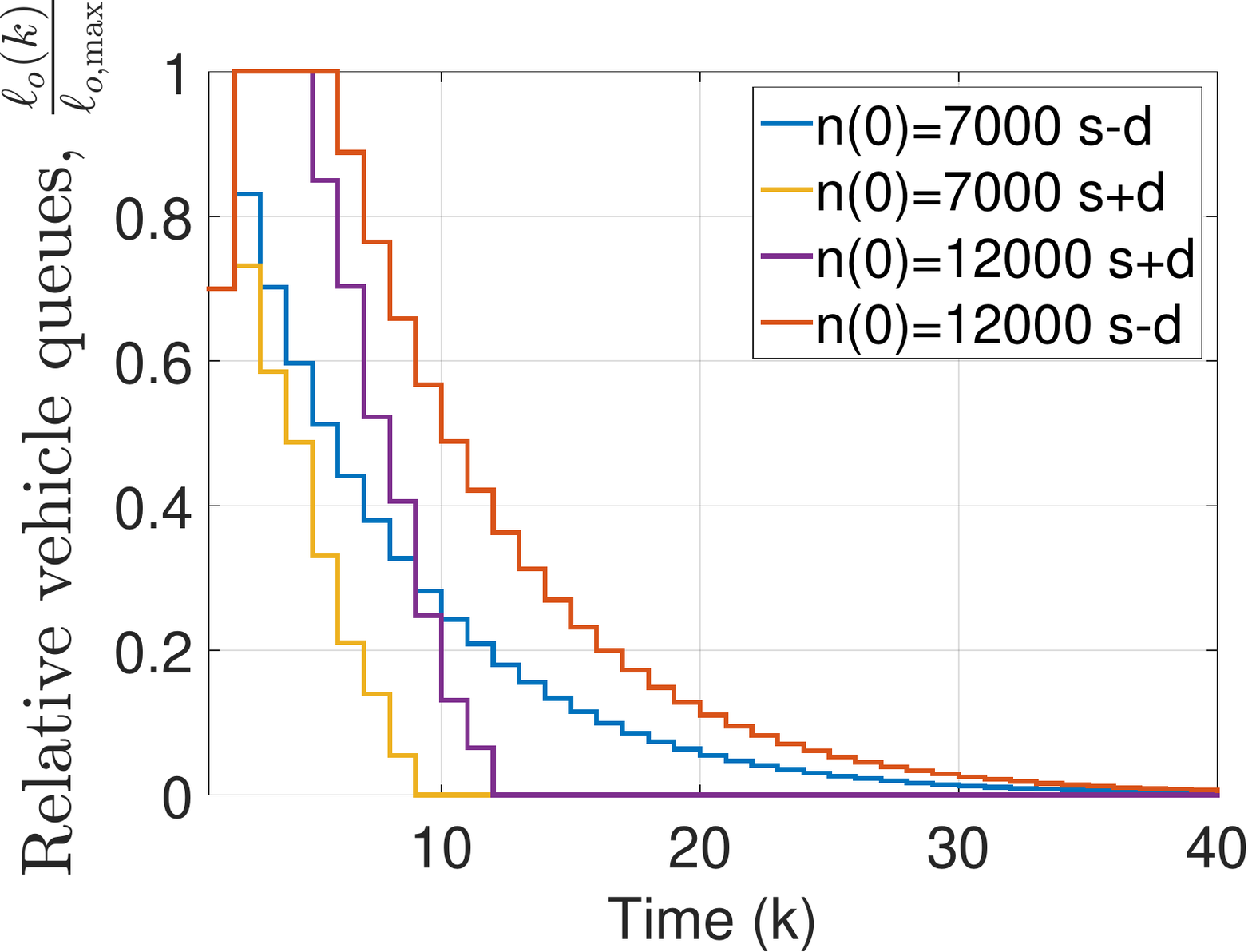}\\
(m) Relative queue: link 8 &     (n) Relative queue: link 9 &     (o) Relative queue: link 13
\end{tabular}
\caption{(a, b)  State and control trajectories without link dynamics; (c) State trajectories of the protected network for different initial points with and without disturbance; (d--o) State and control trajectories of six selected origin links (gates) for different initial points with and without disturbance. For the initial value of vehicle accumulation $n(0)$, see the legend in each subfigure; initial queues at origin links $\ell_o(0) = 0.7 \ell_{o,\max}, \, \forall\, o \in \cal O$.}\label{fig:results}
\end{figure*}

Figures \ref{fig:results}(a) and \ref{fig:results}(b) depict the state and control trajectories for the perimeter flow control problem without origin link dynamics, i.e. for the single-input single output control problem with only state equation \eqref{eq:PNd}. As can be seen, the strategy manages to stabilise the vehicle accumulation of the protected network around its desired point $\hat{n} = 4000$ veh starting from a number of different initial points (including the extreme case of partial gridlock). The strategy restricts flow to enter the protected network area whenever $n > 4000$, while increases the input flows for $n < 4000$. It is evident that the single-region control strategy without queue dynamics outside of the protected area needs less time and effort to stabilise the system at $k=15$, compared to the proposed multi-gated perimeter flow control, which stabilises all queues and protected network's accumulation at $k=20$. This is attributed to the complete lack of information of the geometric characteristics of the origin links that affects control decisions.

\subsection{Performance assessment of proposed gating policies}\label{sec:CompAlloc}

This section compares the proposed multi-gated perimeter flow control strategy (MGC), which explicitly considers the queue dynamics outside the protected network area and operational constraints (minimum/maximum queues, capacities, etc.), with the two perimeter-ordered flow allocation policies presented in Section \ref{sec:allocationpolicies}, namely CAP and  OAP, and the no control case. The storage capacity-based ratios of the fifteen entrance links for CAP are as follows (see Table \ref{tbl:GatesC} for link capacities):   Ratio (\%) =  \{6.2, 5.3, 8.0,  4.8, 4.6,	5.3, 	5.3, 	5.3, 	14.4, 	13.1, 	5.3, 	5.0, 	4.1, 	4.1, 	9.1\}. All strategies were first compared for a \emph{scenario without external demand}, designed to allow for CAP, OAP, and no control to cope with excessive queues at origin links (outside the protected network area) and effectuate a fair comparison. Then two \emph{scenarios with medium and high external demand} were used to demonstrate the equity properties of MGC and its ability to manage excessive queues outside of the protected network area and optimally distribute the input flows with respect to geometric characteristics. Note that controlling the external boundary of a network, restricts vehicles from entering the network resulting in virtual queues, particularly under the no control case. The delays for these vehicles are estimated as they do not have an option to change.

\subsubsection{Comparison between MGC and CAP/OAP} 

\begin{figure*}[tb]\centering
 \begin{tabular}{cc}
\includegraphics[width=.4\textwidth]{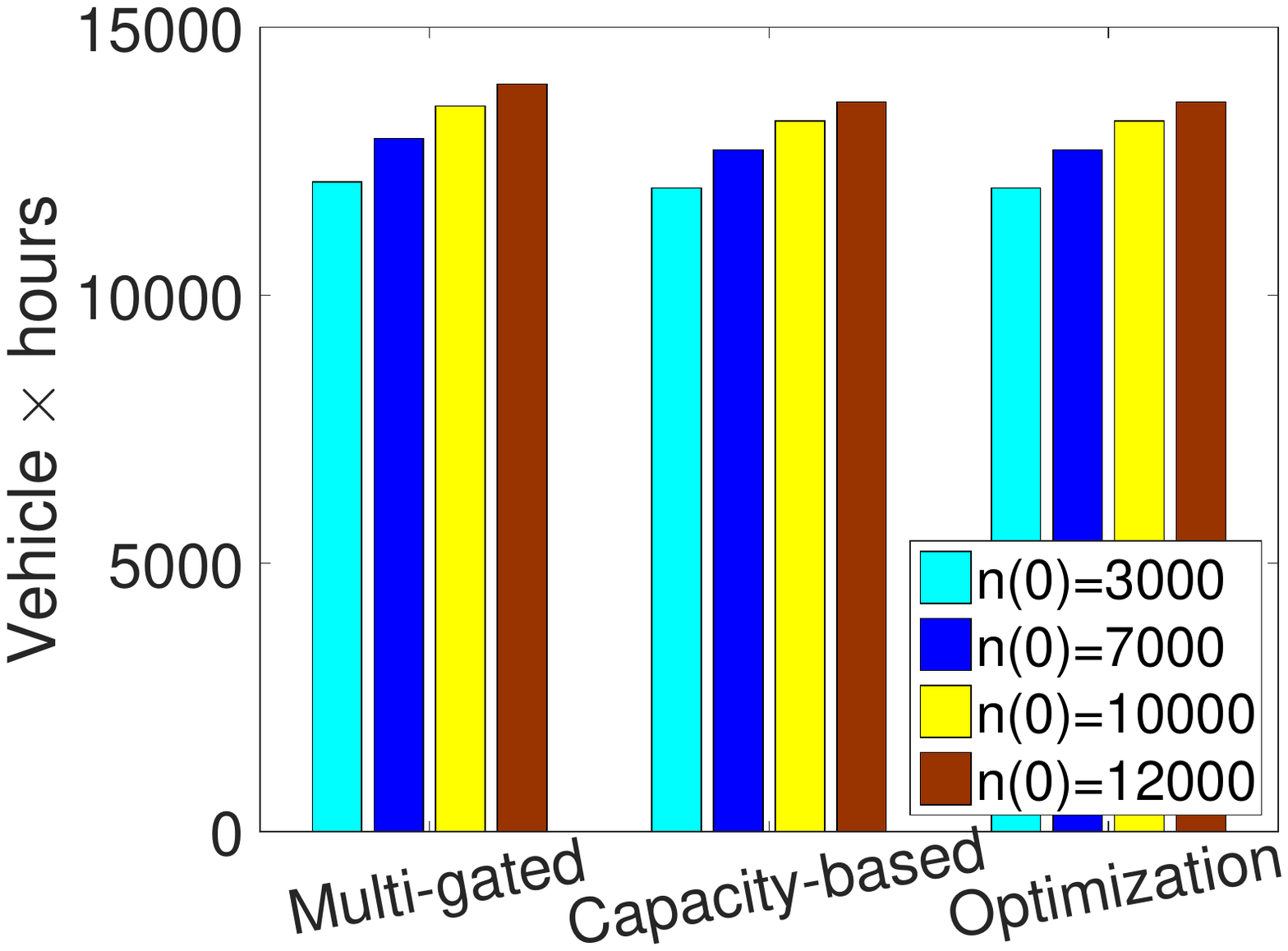} &
 \includegraphics[width=.4\textwidth]{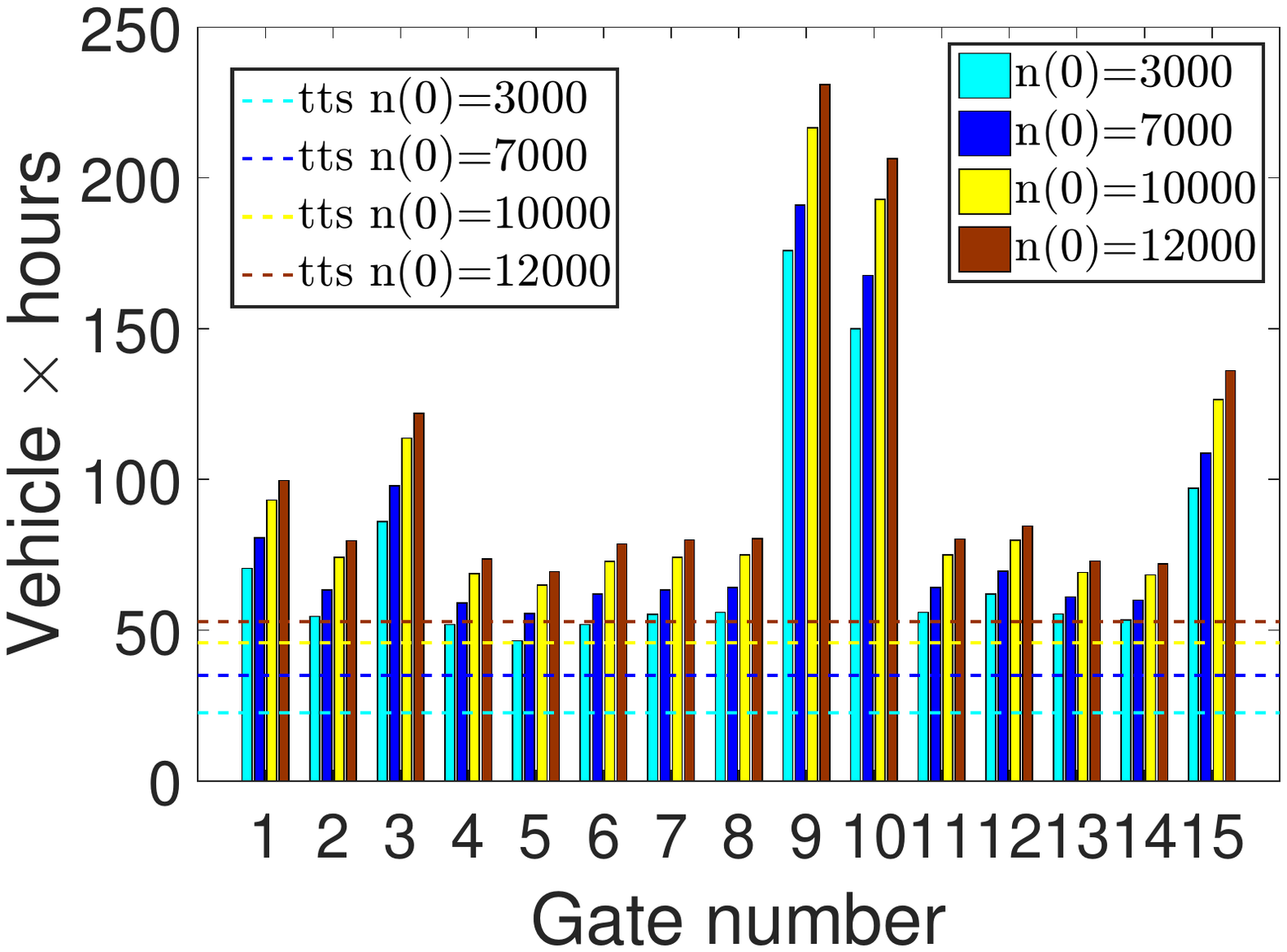}\\
\small{(a) TTS at PN for MGC, CAP, and OAP} & \small{(b)  TTS at gated links with MGC} \\[5pt]
 \includegraphics[width=.4\textwidth]{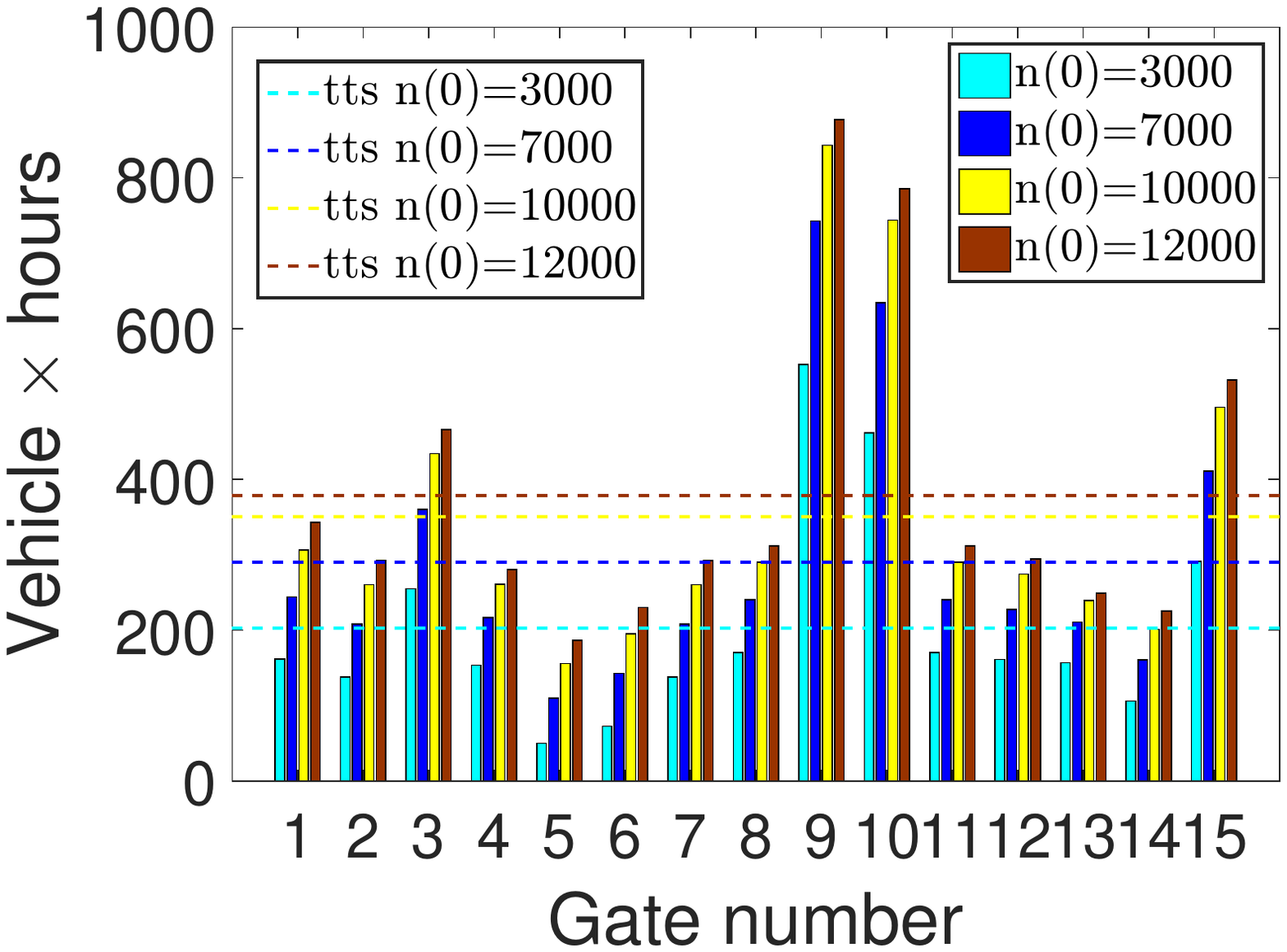} &
 \includegraphics[width=.4\textwidth]{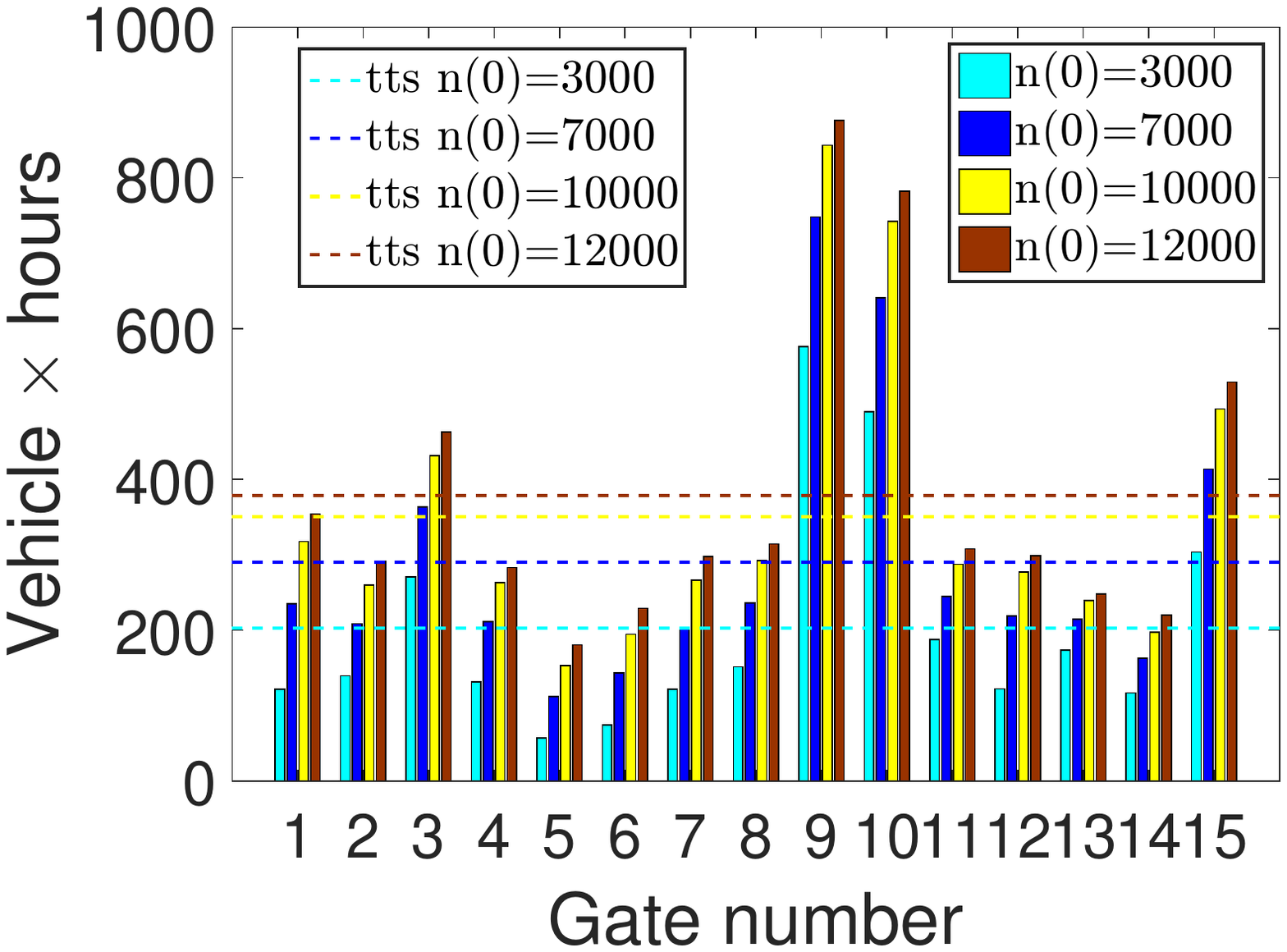}\\
\small{(c) TTS at gated links with CAP} & \small{(d) TTS at gated links with OAP}  \\ 
 \end{tabular}
\caption{Total time spent within the protected network and gated links under MGC, CAP, and OAP for four initial states in a scenario without external demand.}\label{fig:ttsnodemand}
\end{figure*}

Figure\ \ref{fig:ttsnodemand} depicts the obtained results (delays/TTS) within the protected network and gated links under MGC, CAP, and OAP for four initial states in a scenario without external demand. Figure \ref{fig:ttsnodemand}(a) underlines the ability of all strategies to protect the inner network area (PN) of downtown San Francisco from congestion with similar performance across different traffic conditions (ranging from free-flowing to near gridlock). Comparing Figure \ref{fig:ttsnodemand}(b) with Figures \ref{fig:ttsnodemand}(c) and \ref{fig:ttsnodemand}(d), TTS (and delay) under MGC is seen to be four times lower at any gate compared to TTS under OAP/CAP. Table \ref{tbl:Gatesavgtts-d} further supports the above arguments on the superiority of MGC versus OAP/CAP and no control case particularly at the original links outside the protected network. Note that the incoming/gated flows from the periphery of the network is much higher under MGC compared to OAP/CAP. In other words, the MGC strategy serves more cars and maximises throughput for the same simulation horizon. 

\begin{table}[tbp]
\centering
\caption{Average value of TTS (veh $\times$ h) at gated links under No Control, MGC, CAP, and OAP  for different initial states in a scenario without external demand. Initial queues of $\ell_{o}(0)$ at 70\% of $\ell_{o,\max}$ for each origin link.}\label{tbl:Gatesavgtts-d}
\begin{tabular}{l||c|c|c|c}
            \hline%
           Policy/Initial state                         &  No Control & MGC & CAP & OAP\\ \hline           
           $n(0) = 3000$ veh     &  317		& 23	 	& 203   & 203\\	                 
           $n(0) = 7000$ veh     &  415           & 35          & 291   & 291\\
           $n(0) = 10000$ veh   & 480            & 46          & 350   & 351\\              
           $n(0) = 12000$ veh   & 520	        & 53	     	& 378   & 378\\
           \hline                                       
        \end{tabular}
\end{table}

It should be noted that under OAP and CAP drivers experience the same TTS on average, within or outside the protected network area. Though slight differences can be observed from a careful inspection of Figures \ref{fig:ttsnodemand}(c) and \ref{fig:ttsnodemand}(d). This result is attributed to the lack of flexibility (capacity-based metering) of both allocation strategies to manage the developed queues outside the protected network area. Long queues at origin links (without knowledge of queue dynamics and geometric characteristics) result to conservative control and more or less similar releasing rates.

\subsubsection{Equity properties of MGC}

\begin{figure}[tbp]
\centering
\begin{tabular}{c}
 \includegraphics[width=.4\textwidth]{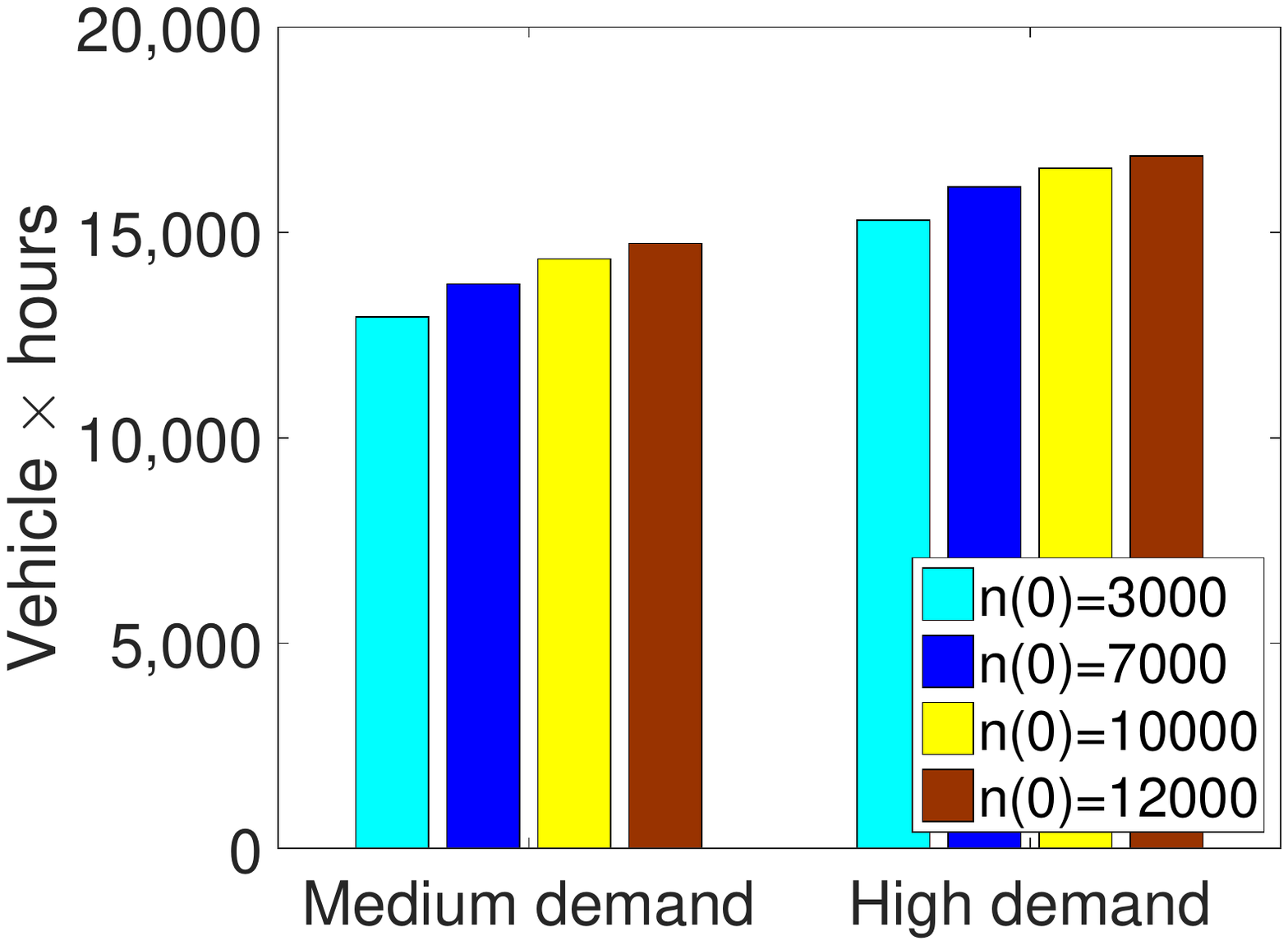}\\
\small{(a) TTS at PN for MGC.} \\[5pt]
\includegraphics[width=.4\textwidth]{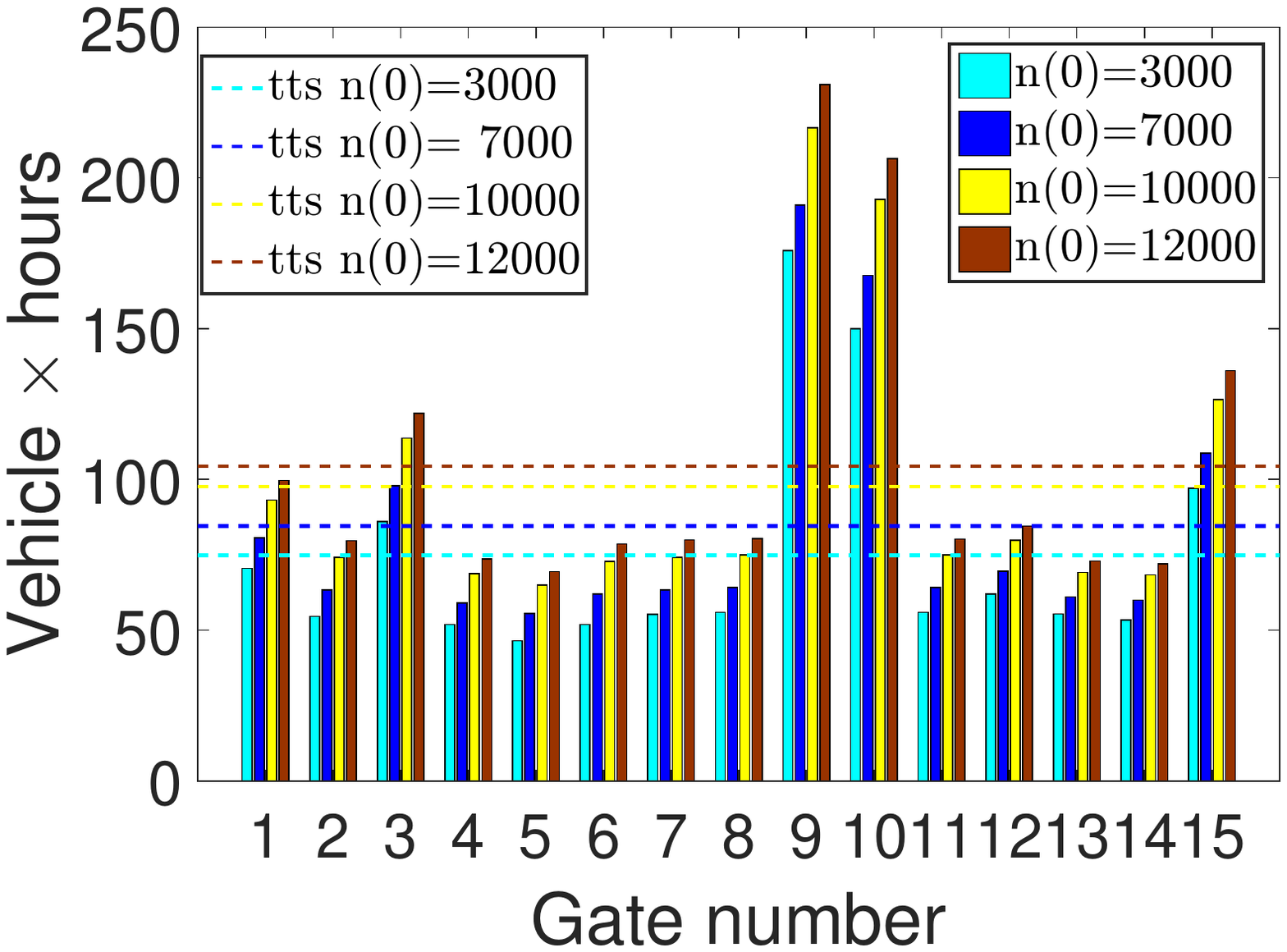} \\
\small{(b) Gated links medium demand.} \\[5pt]
 \includegraphics[width=.4\textwidth]{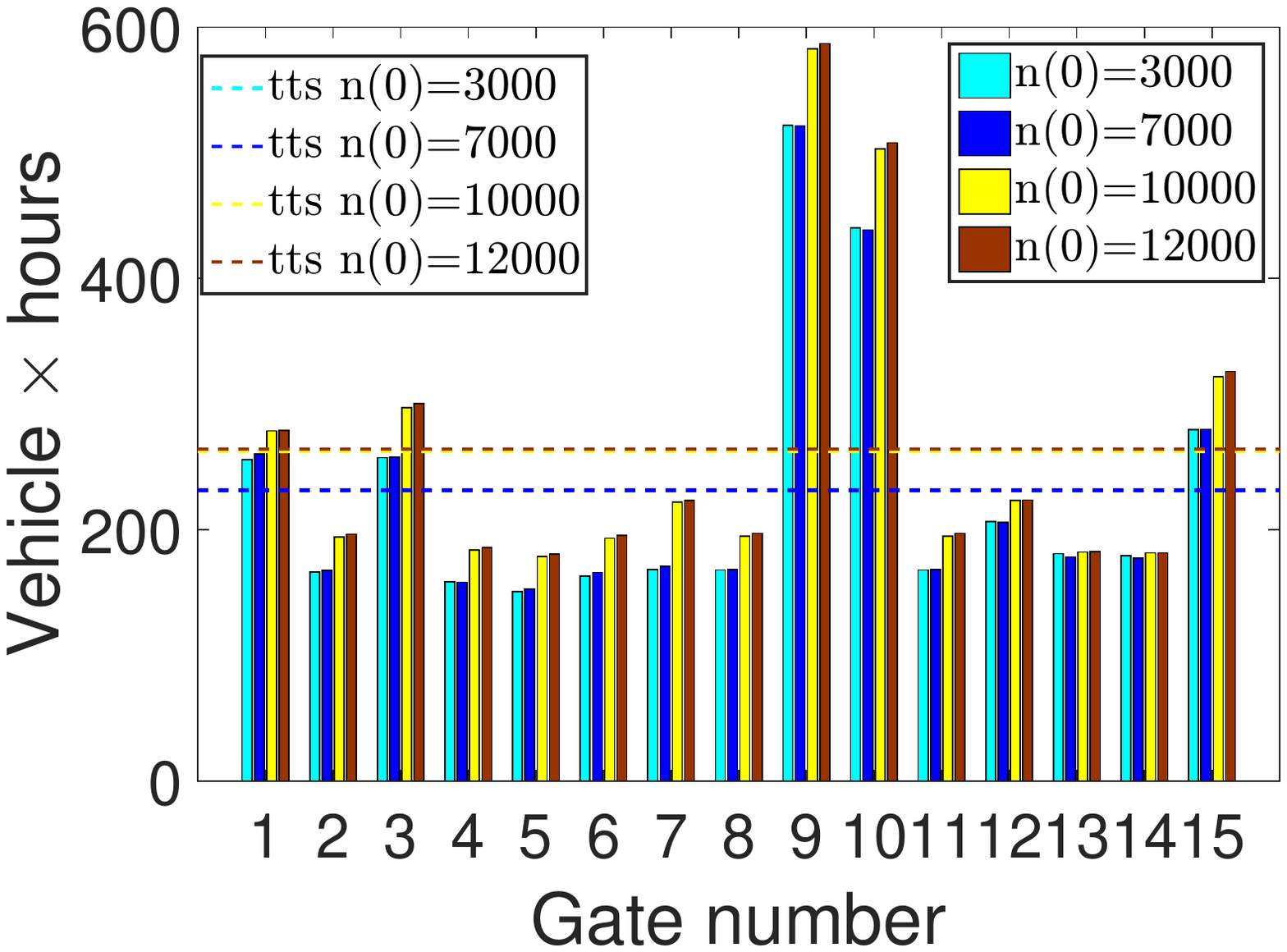}\\
\small{(c) Gated links high demand.} 
\end{tabular}
\caption{Total time spent within the protected network area and gated links under MGC for different initial states and two scenarios with external demand.}\label{fig:ttsd}
\end{figure}

Figure\ \ref{fig:ttsd} demonstrates the equity properties of the proposed MGC approach to better manage excessive queues outside of the protected network area and optimally distribute the input flows with respect to geometric characteristics. These figures depict results obtained for two different demand scenarios, namely medium and high, and four initial states $n(0)$ (congested and semi-congested regimes) of the fundamental diagram. As can be seen in Figure\ \ref{fig:ttsd}(a), Total Time Spent (TTS) within the protected network area increases with vehicle accumulation for both demand scenarios. Remarkably, Figures \ref{fig:ttsd}(b)--\ref{fig:ttsd}(c)  demonstrate the equity properties of the proposed multi-gated perimeter flow control. More precisely, gates with similar geometric characteristics experience similar TTS (delays) for two different demand scenarios and four different initial states in the protected network area. For instance, we can distinguish three different groups of gates with similar TTS, Group A including gates 2, 4--8, 11--14; Group B including gates 1, 3, 15; and Group C including gates 9, 10. Contrasting gates in Groups A, B, and C with the geometric characteristics in Table \ref{tbl:GatesC}, further supports the equity properties of the proposed multi-gated control. It should be noted that delays incurred at gated links under MGC for the high demand scenario are comparable or lower (for saturated conditions) to those obtained under CAP and OAP policies under the medium demand scenario. Note that CAP and OAP policies, and the no control case were not be able to cope with the high demand scenario (both incurred excessive queues and delays near gridlock), which is attributed to the lack of knowledge of the traffic dynamics outside of the protected network area (no entrance link dynamics).

Concluding, the control flexibility and efficiency of the proposed control (for a number of performance assessment criteria -- qualitative and quantitative), while explicitly considering the queue dynamics and constraints, underlines the clear superiority of appropriate multi-gated perimeter flow control. Certainly, it should be highlighted the efficiency and equity properties of the proposed multi-gated scheme to better manage excessive queues outside of the protected network area and optimally distribute the input flows.

\section{Conclusions and outlook}\label{sec:conclusions}

In this paper, an integrated model for multi-gated perimeter flow control is presented. Compared to previous works, the proposed scheme determines optimally distributed input flows for a number of gates located at the periphery of a protected network area. 
We also proposed practical  flow allocation policies without entrance link dynamics (without constraints), namely, capacity-based flow allocation policy (CAP) and optimisation-based flow allocation policy (OAP).

Simulation results for a protected  area of downtown San Francisco with fifteen gates of different geometric characteristics were presented. Results demonstrated the efficiency and equity properties of the proposed MGC approach to better manage excessive queues outside of the protected network area and optimally distribute the input flows compared to single-region perimeter flow control (without queue dynamics). It is expected that similar policies can also be  utilised for dynamic road pricing. 

The flow allocation policies, CAP and OAP, indicated more or less the same performance for the considered simulation scenarios, where both demonstrated slow performance in dissolving the developed queues at the origin links. The TTS performance for MGC observed to be profoundly lower compared to TTS of both flow allocation policies. This underlined that MGC strategy serves more cars and maximises throughput for the same simulation horizon. By considering the queue dynamics and constraints underlines the clear superiority of appropriate MGC-based schemes.

The equity properties of perimeter flow control have not attracted considerable attention in the literature, although it is an important characteristic of any practical perimeter flow control application. Future research will focus on: (a) extending the proposed framework for multi-region cities described by three-dimensional bi-modal passenger and vehicle MFDs; (b) the efficiency versus equity properties of perimeter flow control with queue dynamics, and; (c) the integration of additional components for dynamic routing and road pricing.

\section*{Acknowledgements}
The first author would like to acknowledge support by the Malaysian Government and the National Defense University of Malaysia (NDUM). The second author would like to acknowledge support by the Economic and Social Research Council  (ESRC), UK Research and Innovation (UKRI) (grant number ES/L011921/1). 

%


\printcredits

\bibliographystyle{cas-model2-names}

\bibliography{MyBibliography}

%

\end{document}